\documentclass[10pt,letterpaper]{article}
\pdfoutput=1

\usepackage{jcappub}
\usepackage{amsmath,amssymb,color}
\usepackage{enumerate,xstring}
\usepackage{jcappub}
\usepackage{amsmath}
\usepackage{amsfonts}
\usepackage{amssymb}
\usepackage{graphicx}
\usepackage{epsfig}
\usepackage{color}
\usepackage{multirow}
\usepackage{graphicx}
\usepackage{bigints}
\usepackage{todonotes,bm,etoolbox}
\usepackage{calligra}
\usepackage{hyperref}
\usepackage{tabularx}
\usepackage{booktabs}
\usepackage{subfigure}

\usepackage{tikz}

\def\ba#1\ea{\begin{align}#1\end{align}}
\def\bea{\begin{eqnarray}}
\def\eea{\end{eqnarray}}
\def\be{\begin{equation}}
\def\ee{\end{equation}}

\def\({\left(}
\def\){\right)}
\def\[{\left[}
\def\]{\right]}

\def\<{\left\langle}
\def\>{\right\rangle}

\def\comment#1{}

\def\eps{\epsilon}

\renewcommand{\v}[1]{\bm{#1}}

\def\vx{\v{x}}
\def\vk{\v{k}}
\def\vq{{\v{q}}}

\def\vtheta{\v{\theta}}

\def\vD{\v{D}}

\newcommand{\perm}[1]{ \expandafter\ifstrempty\expandafter{#1} {\mbox{perm.}} {\mbox{$#1$ perm.}} }

\DeclareMathOperator{\Cov}{\rm \bf Cov}
\DeclareMathOperator{\cov}{\rm \bf Cov}

\def\P{\mathcal{P}}

\newcommand{\fnl}{f_\textnormal{\textsc{nl}}}
\newcommand{\bphi}{b_\phi}
\newcommand{\bphidelta}{b_{\phi\delta}}
\newcommand{\A}{\mathcal{A}}
\newcommand{\bigbox}{L_{\rm box}\approx 800\ {\rm Mpc}}

\definecolor{RedWine}{rgb}{0.743,0,0}
\definecolor{RoyalBlue}{rgb}{0.25,.41,.88}
\definecolor{ForestGreen}{rgb}{.13,.54,.13}
\definecolor{Goldenrod}{rgb}{.85,.65,.13}

\newcommand{\bq}{\begin{eqnarray}}
\newcommand{\eq}{\end{eqnarray}}

\title{\huge Predictions for local PNG bias in the galaxy power spectrum and bispectrum and the consequences for $\fnl$ constraints}

\author[a,b]{Alexandre Barreira}

\affiliation[a]{\small Excellence Cluster ORIGINS, Boltzmannstra\ss e 2, 85748 Garching, Germany}
\affiliation[b]{\small Ludwig-Maximilians-Universit\"at, Schellingstra\ss e 4, 80799 M\"unchen, Germany}

\emailAdd{alex.barreira@origins-cluster.de}

\date{\today}

\abstract{We use hydrodynamical separate universe simulations with the IllustrisTNG model to predict the local primordial non-Gaussianity (PNG) bias parameters $\bphi$ and $\bphidelta$, which enter at leading order in the galaxy power spectrum and bispectrum. This is the first time that $\bphidelta$ is measured from either gravity-only or galaxy formation simulations. For dark matter halos, the popular assumption of universality overpredicts the $\bphidelta(b_1)$ relation in the range $1 \lesssim b_1 \lesssim 3$ by up to $\Delta\bphidelta \sim 3$ ($b_1$ is the linear density bias). The adequacy of the universality relation is worse for the simulated galaxies, with the relations $\bphi(b_1)$ and $\bphidelta(b_1)$ being generically redshift-dependent and very sensitive to how galaxies are selected (we test total, stellar and black hole mass, black hole mass accretion rate and color). The uncertainties on $\bphi$ and $\bphidelta$ have a direct, often overlooked impact on the constraints of the local PNG parameter $\fnl$, which we study and discuss. For a survey with $V = 100{\rm Gpc}^3/h^3$ at $z=1$, uncertainties $\Delta\bphi \lesssim 1$ and $\Delta\bphidelta \lesssim 5$ around values close to the fiducial can yield relatively unbiased constraints on $\fnl$ using power spectrum and bispectrum data. We also show why priors on galaxy bias are useful even in analyses that fit for products $\fnl\bphi$ and $\fnl\bphidelta$. The strategies we discuss to deal with galaxy bias uncertainties can be straightforwardly implemented in existing $\fnl$ constraint analyses (we provide fits for some of the bias relations). Our results motivate more works with galaxy formation simulations to refine our understanding of $\bphi$ and $\bphidelta$ towards improved constraints on $\fnl$.}

\begin{document}

\maketitle

\section{Introduction}
\label{sec:intro}

One of the main open questions in cosmology today concerns the origin of the primordial density fluctuations generated in the early Universe during the epoch of inflation. The simplest theoretical explanation is that of single-field models, in which the density perturbations are due to quantum fluctuations of a single scalar degree of freedom that rolls slowly down its potential. These models crucially predict density fluctuations that are Gaussian distributed \cite{maldacena:2003, 2004JCAP...10..006C, 2011JCAP...11..038C, Tanaka:2011aj, conformalfermi, 2015JCAP...10..024D}, and as a result, the detection of any deviation from Gaussianity would immediately rule them out and open the door for more elaborate, multifield models \cite{2014arXiv1412.4671A, 2019Galax...7...71B}. {\it Local-type} primordial non-Gaussianity (PNG) is the most popular way to describe departures from perfectly Gaussian-distributed fluctuations. In this case, the primordial gravitational (Bardeen) potential $\phi(\vx)$ is expanded as \cite{2001PhRvD..63f3002K}
\bq\label{eq:fnl}
\phi(\vx) = \phi_{\rm G}(\vx) + \fnl\left[\phi_{\rm G}(\vx)^2 - \left<\phi_{\rm G}(\vx)^2\right>\right],
\eq
where $\phi_{\rm G}$ is a Gaussian distributed random field and $\left<\cdots\right>$ indicates ensemble averaging. The parameter $\fnl$ is a constant that quantifies the leading-order departure from Gaussianity (this expansion can continue to include third and higher powers of $\phi_{\rm G}$). Analyses of the three-point function of the cosmic microwave background (CMB) radiation measured by the Planck satelllite currently constrain $\fnl = -0.9 \pm 5.1\ (1\sigma)$ \cite{2019arXiv190505697P}, but next-generation large-scale structure surveys hope to improve upon this bound.

As galaxies form out of the primordial fluctuations, the statistics of their distribution can in principle be used to probe PNG. In cosmologies with local PNG, the deterministic galaxy density contrast $\delta_g$ can be written to leading order (LO) as  \cite{mcdonald:2008, giannantonio/porciani:2010, assassi/baumann/schmidt}
\bq\label{eq:delta_g_LO}
\delta_g(\vx, z) \stackrel{\rm LO}{=} b_1(z)\delta_m(\vx, z) + \bphi(z) \fnl\phi(\vq),
\eq
where $\delta_m$ is the total matter density contrast and $\vq$ is the initial Lagrangian coordinate associated with the final Eulerian coordinate $\vx$. In keeping with the effective field theory approach to galaxy statistics and the galaxy bias expansion (see Ref.~\cite{biasreview} for a comprehensive review), the wavelength of the perturbations $\delta_m(\vx, z)$ and $\phi(\vq)$ is implicitly assumed to be much larger than the spatial scales over which galaxy formation takes place. The coefficients $b_1$ and $\bphi$ are called galaxy bias parameters and they describe, respectively, the linear {\it responses} of the galaxy number density to total mass and primordial potential perturbations with $\fnl \neq 0$. The bias parameters effectively encode all of the complicated dependence of galaxy formation on the large-scale environment, are functions not only of redshift but also of the properties of the galaxies considered, and are thus extremely challenging to predict theoretically. The bias parameters are therefore normally fitted alongside the cosmological ones in inference analyses using galaxy clustering data, but as degeneracies between the two sets of parameters arise, one often finds it necessary/beneficial to take theoretical priors on bias into account in order to tighten the constraints on cosmology. These degeneracies are especially critical in $\fnl$ constraints. For example, Ref.~\cite{dalal/etal:2008} showed that the galaxy power spectrum $P_{gg}(k)$ (the Fourier transform of the two-point correlation function) in cosmologies with local PNG gets a contribution $\propto b_\phi \fnl / k^2$, that becomes important on large scales and that can be used to constrain $\fnl$. Given the perfect degeneracy between $\fnl$ and $\bphi$, however, it is simply impossible to use $P_{gg}$ to constrain $\fnl$, unless some theoretical prior on $\bphi$ is assumed\footnote{Formally, next-to-leading-order contributions to the galaxy power spectrum (e.g.~1-loop terms and beyond) can break the degeneracy, but they are unimportant given the current observationally allowed range for $\fnl$.}.

In the large-scale structure literature, the most popular way to circumvent this problem in $\fnl$ constraints is by relating $\bphi$ to $b_1$ using the so-called {\it universality relation} \cite{dalal/etal:2008, slosar/etal:2008, 2008ApJ...677L..77M, giannantonio/porciani:2010}
\bq\label{eq:bphi_uni}
b_\phi = 2\delta_c\left(b_1 - 1\right),
\eq
which follows from assuming that the halo mass function is universal, and $\delta_c = 1.686$ is the threshold overdensity for spherical collapse. In this way, the contribution scales as $\propto\left(b_1-1\right)\fnl/k^2$, and since $b_1$ can in principle be constrained using the smaller-scale part of the power spectrum (where $\fnl$ contributes weakly), it then becomes possible to constrain $\fnl$. The universality relation is adopted by almost all existing galaxy data constraints on $\fnl$ \cite{slosar/etal:2008, 2011JCAP...08..033X, 2013MNRAS.428.1116R, 2014PhRvD..89b3511G, 2014PhRvL.113v1301L, 2014MNRAS.441L..16G, 2015JCAP...05..040H, 2019JCAP...09..010C, 2021arXiv210613725M} (the current tightest bound is $\fnl = -12 \pm 21\ (1\sigma)$ \cite{2021arXiv210613725M}), as well as in forecast studies \cite{2012MNRAS.422.2854G, 2014arXiv1412.4872D, 2014arXiv1412.4671A, 2015ApJ...814..145A, 2015PhRvD..92f3525A, 2016JCAP...05..009R, 2017PhRvD..95l3513D, 2018MNRAS.478.1341K, 2019ApJ...872..126M, 2019MNRAS.489.1950B, 2021JCAP...04..013M, 2021arXiv210609713S} for next-generation surveys. Despite its widespread adoption, there is however no reason to expect the universality relation to hold for real-life galaxy samples, and in fact, studies using $N$-body simulations have been indicating this to be the case already. For example, gravity-only simulations have shown that the $\bphi(b_1)$ relation of dark matter halos underpredicts slightly the universality relation \cite{grossi/etal:2009, desjacques/seljak/iliev:2009, 2010MNRAS.402..191P, 2011PhRvD..84h3509H, 2017MNRAS.468.3277B}, Ref.~\cite{2020JCAP...12..013B} showed using galaxy formation simulations that the $\bphi(b_1)$ relation of stellar-mass selected galaxies overpredicts the universality relation, and Refs.~\cite{slosar/etal:2008, 2010JCAP...07..013R} discussed how the $\bphi(b_1)$ relation of recently-merged objects differs also from the universality prediction; we will return to these findings below when we reproduce some with our numerical results. The impact that uncertainties on the $\bphi(b_1)$ relation have on the resulting $\fnl$ constraints is only now beginning to be explored \cite{2020JCAP...12..031B, 2021JCAP...05..015M}, but it is crucial that these investigations are made robust and mature since they will ultimately directly impact the final constraints on $\fnl$.

Beyond leading-order, the contributions at next-to-leading-order (NLO) are\footnote{There is an additional contribution $\delta_g \supset b_{\phi^2}\fnl^2\phi^2$ that is quadratic in $\fnl$, but which is unimportant given the current observational bounds, and so we skip writing it explicitly. Should this contribution have been important, then all our motivations to study $\bphidelta$ in this paper would apply equally to $b_{\phi^2}$.}
\bq\label{eq:delta_g_MLO}
\delta_g(\vx, z) \stackrel{\rm NLO}{=} \frac{b_2(z)}{2}\left[\delta_m(\vx, z)\right]^2 + b_{K^2}(z)\left[K_{ij}(\vx, z)\right]^2 + \bphidelta(z)\fnl\phi(\vq)\delta_m(\vx, z),
\eq
where $K_{ij} = \big(\partial_i\partial_j/\nabla^2 - \delta_{ij}/3\big)\delta_m$ is a long-wavelength tidal field, and $b_2$, $b_{K^2}$ and $\bphidelta$ are additional galaxy bias parameters. These terms contribute to the galaxy bispectrum $B_{ggg}(k_1,k_2,k_3)$ (the Fourier transform of the three-point correlation function), which is also a very well-known probe of $\fnl$ \cite{scoccimarro/etal:2004, 2007PhRvD..76h3004S, jeong/komatsu:2009b, 2011JCAP...04..006B, sefusatti/etal:2012}, and which may prove crucial to combine with the galaxy power spectrum if next-generation galaxy surveys are to improve over the current CMB constraints. For halos, the $b_2(b_1)$ relation is very well understood \cite{lazeyras/etal} and works have been progressively improving our understanding of $b_{K^2}(b_1)$ as well \cite{sheth/chan/scoccimarro:2012, baldauf/etal:2012, 2018JCAP...09..008L, saito/etal:14, 2018JCAP...07..029A, 2021arXiv210206902E, 2021arXiv210502876B, 2021arXiv210614713L}; the corresponding relations for simulated galaxies have also began to be recently studied \cite{2021arXiv210502876B}. In contrast, the $\bphidelta$ parameter has never been the focus of dedicated simulation work. Assuming universality of the halo mass function, one finds \cite{giannantonio/porciani:2010, biasreview}
\bq\label{eq:bphidelta_uni}
\bphidelta = b_\phi - b_1 + 1 + \delta_c[b_2 - (8/21)(b_1-1)],
\eq
but unlike the case for $\bphi$, the performance of this relation has never been checked, even for the simpler case of dark matter halos in gravity-only simulations. There are currently no real-data constraints on $\fnl$ using the galaxy bispectrum, but existing forecast studies effectively always assume the validity of this relation. Although it is formally possible to constrain $\fnl$ using the bispectrum without any prior on $\bphidelta$, we will see below that adopting priors based on a $\bphidelta(b_1)$ relation is necessary for competitive constraints. 

This motivates the main goal of this paper, which is to use separate universe $N$-body simulations to obtain predictions for the bias parameters $\bphi$ and $\bphidelta$; to the best of our knowledge, this is the first time that $\bphidelta$ is estimated from dedicated simulation data. We will do so not only for the case of dark matter halos in gravity-only simulations, but also for simulated galaxies in hydrodynamical simulations with the IllustrisTNG galaxy formation model. One of our main new results is that the universality relation of Eq.~(\ref{eq:bphidelta_uni}) is not a perfect description for either one of these large-scale structure tracers, with the size of the departures varying depending on the galaxy selection criteria adopted. Recently, Ref.~\cite{2021arXiv210502876B} showed that the $b_2(b_1)$ and $b_{K^2}(b_1)$ relations of halos are broadly preserved for galaxies selected by a variety of criteria (total mass, stellar mass, color and black hole mass accretion rate), which suggests that priors around these relations may be adopted relatively safely in real-data analyses. In contrast, our results below will show that the $\bphi(b_1)$ and $\bphidelta(b_1)$ relations are markedly more sensitive to the selection criteria, which makes the design of theoretical priors more challenging. In a second part of this paper, we build upon the idealized forecast study of Ref.~\cite{2020JCAP...12..031B} to study the impact that uncertainties on $\bphi$ and $\bphidelta$ can have on $\fnl$ constraints. We will see that these bias parameters can have a marked impact on the resulting $\fnl$ bounds. One of our main takeaway messages is that more simulation work is needed to our current level 


The rest of this paper is organized as follows. In Sec.~\ref{sec:method} we describe the simulation data we use in this work, as well as our methods to estimate the galaxy bias parameters. Section \ref{sec:results} contains our main numerical results on galaxy bias, where we show first the $\bphidelta$ parameter measured in gravity-only simulations, and then compare the sensitivity of the $\bphi(b_1)$ and $\bphidelta(b_1)$ relations to different galaxy selection criteria in hydrodynamical simulations. In Sec.~\ref{sec:forecasts} we show and discuss the results of a simple forecast study for a fictitious survey aiming to illustrate the impact that bias uncertainties have on $\fnl$ constraints. We summarize and conclude in Sec.~\ref{sec:summary}. In App.~\ref{app:b2}, we show estimates of the bias parameter $b_2$ using the same method we use to estimate $\bphidelta$. In App.~\ref{app:theory}, we collect the theory model expressions for the galaxy power spectrum and bispectrum that we use in our forecast study.

\section{Methodology}
\label{sec:method}

In this section we describe the estimation of the two leading-order local PNG galaxy bias parameters $\bphi$ and $\bphidelta$, as well as the linear bias parameter $b_1$, which recall, contribute to the bias expansion as
\bq\label{eq:biasexp_set}
\delta_g(\vx, z) \supset b_1(z)\delta_m(\vx, z) + \bphi(z)\fnl\phi(\vq) + \bphidelta(z)\fnl\phi(\vq)\delta_m(\vx, z).
\eq
As explained below, we estimate (i) $b_1$ using the large-scale limit of the galaxy-matter cross-power spectrum; (ii) $b_\phi$ as the response of the galaxy number density to long-wavelength primordial gravitational potentials with $\fnl$; and (iii) $b_{\phi\delta}$ using the response of $b_1$ to long-wavelength primordial gravitational potentials with $\fnl$.

\subsection{Simulation data specifications}
\label{sec:method_sepuni}

The simulations we use in this work have been presented previously in Ref.~\cite{2020JCAP...12..013B}, and they were run using the moving-mesh gravity+hydrodynamical $N$-body code {\sc AREPO} \citep{2010MNRAS.401..791S, 2016MNRAS.455.1134P} with the IllustrisTNG model of galaxy formation \citep{2017MNRAS.465.3291W, Pillepich:2017jle,Nelson:2018uso}. This model is an improved version of its precursor Illustris \cite{2014MNRAS.445..175G, 2014MNRAS.444.1518V}, and it is characterized by {\it sub-grid} prescriptions for the physics of gas cooling, star formation, stellar feedback, chemical enrichment, and black hole growth/feedback, which were calibrated to broadly reproduce a number of observations such as the galaxy stellar mass function at low redshift, the star formation rate history, galaxy sizes and the gas mass fractions of galaxies and galaxy groups (see Refs.~\cite{2018MNRAS.480.5113M, Pillepich:2017fcc, 2018MNRAS.477.1206N, 2018MNRAS.475..676S, Nelson:2017cxy, 2019MNRAS.490.3234N, 2019MNRAS.490.3196P} for the first results with IllustrisTNG). The initial conditions were generated at  $z_i = 127$ with {\sc N-GenIC} code \citep{2015ascl.soft02003S} using the Zel'dovich approximation, and a linear matter power spectrum calculated using the {\sc CAMB} code \citep{camb}. The galaxy formation simulations were run on a cubic box with size $L_{\rm box} = 205\ {\rm Mpc}/h$, containing $N_p = 1250^3$ dark matter mass elements and $N_p = 1250^3$ initial gas elements. At this resolution, we have both full-physics hydrodynamical simulations, as well as gravity-only counterparts (we label these as ``Gravity''); we follow the standard IllustrisTNG nomenclature and refer to this resolution as TNG300-2. In addition, we consider also a set of gravity-only simulations run also with {\sc AREPO} with $N_p = 1250^3$ dark matter mass elements, but on a bigger simulation box size $L_{\rm box} = 560\ {\rm Mpc}/h \approx 800\ {\rm Mpc}$. 

The cosmological parameters of our Fiducial cosmology are: mean baryon density today $\Omega_{b0} = 0.0486$, mean total matter density today $\Omega_{m0} = 0.3089$, mean dark energy density today $\Omega_{\Lambda0} = 0.6911$, dimensionless Hubble rate $h = 0.6774$, primordial scalar spectral index $n_s = 0.967$, and primordial scalar power spectrum amplitude $\A_s = 2.068 \times 10^{-9}$ (at $k_{\rm pivot} = 0.05/{\rm Mpc}$, corresponding to $\sigma_8(z=0) = 0.816$). In order to measure the galaxy bias parameters $\bphi$ and $\bphidelta$ using the separate universe method (see below), we consider also two additional cosmologies, dubbed High$\A_s$ and Low$\A_s$, which differ from the Fiducial only in the value of $\A_s \to \A_s\left[1 + \delta\A_s\right]$, where $\delta\A_s = +0.05$ for High$\A_s$ and $\delta\A_s = -0.05$ for Low$\A_s$. 

We will show measurements of the bias parameters for both halos and subhalos/galaxies. The halos are identified with a Friends-of-Friends (FoF) algorithm run on the dark matter elements with linking length $b = 0.2$ times the mean interparticle distance. In turn, the subhalos correspond to the gravitationally bound structures found by the {\sc SUBFIND} algorithm \cite{2001MNRAS.328..726S} inside each halo. In the hydrodynamical simulations, we refer to the subhalos that contain any mass in stars ($M_* > 0$) as galaxies, and we do not explicitly distinguish between main and satellite galaxies. We also only consider objects with at least $100$ member star particles to ensure we deal with objects that are sufficiently well resolved in the simulations. Throughout, we will show results for objects selected by their total mass $M_t$, stellar mass $M_*$, black hole mass $M_{\rm BH}$, black hole mass accretion rate $\dot{M}_{\rm BH}$, and (dust-uncorrected) $g-r$ color. When quoting the value of these quantities for a given object (halo or subhalo), we always consider the summed contribution of all member particles to that quantity. 

There are two points worth emphasizing about our numerical setup. One is that the TNG300-2 resolution is below that at which the IllustrisTNG model was callibrated at, and as a result, the predictions of our simulations are not in as good agreement with the above-mentioned observations. The galaxy bias predictions are however expected to be less affected by numerical resolution compared to quantities like the galaxy number density itself. In fact, Ref.~\cite{2020JCAP...12..013B} has shown that the $b_1$ and $\bphi$ predictions at TNG300-2 resolution are in very good agreement with those obtained at a higher resolution ($N_p = 2\times1250^3$, $L_{\rm box} = 75{\rm Mpc}/h$; called TNG100-1.5 there) that is closer to the nominal IllustrisTNG one ($N_p = 2\times1820^3$, $L_{\rm box} = 75{\rm Mpc}/h$). The second point is that (as in past works with separate universe simulations of galaxy formation \cite{2019MNRAS.488.2079B, 2020JCAP...02..005B, 2020JCAP...12..013B}) we keep the parameters of the IllustrisTNG model fixed when we adjust the parameter $\A_s$ in the High$\A_s$ and Low$\A_s$ cosmologies. This is the appropriate choice in order to interpret our galaxy bias measurements as predictions of the IllustrisTNG model, i.e., the galaxy bias parameters are the response of galaxy formation to long-wavelength perturbations, at fixed galaxy formation prescription (that of IllustrisTNG). 

\subsection{The linear density and local PNG bias parameters $b_1$ and $b_\phi$}
\label{sec:method_bphi}

We estimate the linear bias parameter $b_1$ in the Fiducial simulations using the large-scale limit of the ratio of the galaxy-matter cross-power spectrum $P_{gm}(k,z)$ and matter power spectrum $P_{mm}(k,z)$
\bq\label{eq:b1_method}
b_1 = \lim_{k \to 0} \frac{P_{gm}(k)}{P_{mm}(k)},
\eq
which follows from Eq.~(\ref{eq:biasexp_set}) for $\fnl = 0$ (note we dropped the redshift $z$ from the arguments to lighten the notation). For the relatively small volume of the TNG300-2 simulations, the scale-dependence of this ratio can still be nonnegligible on the largest scales probed. Thus, rather than simply fitting a constant to it, we account for the leading-order scale-dependence by fitting instead for $b_1 + Ak^2$, and take the constant coefficient as our estimate of the bias parameter. In practice, we use all modes with $k < 0.15h/{\rm Mpc}$, and we have checked that the values of $b_1$ estimated from the TNG300-2 box agree with those from the $\bigbox$ box that is less affected by this systematic. Our error bars on $b_1$ are the error estimate from the least-squares fitting procedure.

On the other hand, we estimate the linear local PNG bias parameter $\bphi$ from the definition
\bq\label{eq:bphi_method}
\bphi = \frac{1}{\bar{n}_g} \frac{\partial \bar{n}_g}{\partial(\fnl\phi)} \equiv \frac{4}{\bar{n}_g} \frac{\partial \bar{n}_g}{\partial\delta\A_s},
\eq
where the first equality follows from Eq.~(\ref{eq:biasexp_set}) and the second equality from the separate universe (or peak-background split) equivalence between (i) structure formation inside a long-wavelength potential perturbation $\fnl\phi$ and (ii) structure formation without the perturbation, but with a modified value of the primordial scalar power spectrum amplitude $\A_s$  \cite{dalal/etal:2008, slosar/etal:2008}. In particular, it is possible to show (see e.g.~Sec.~7.1.2.~of Ref.~\cite{biasreview}) that if $\phi_L$ is the amplitude of a long-wavelength potential perturbation, then galaxies forming inside it form as they would form in a cosmology with $\A_s$ rescaled as $\A_s \to \A_s\left[1 + \delta\A_s\right]$, with $\delta\A_s = 4\fnl\phi_L$ (hence the factor of $4$ in Eq.~(\ref{eq:bphi_method})). Concretely, we evaluate Eq.~(\ref{eq:bphi_method}) via finite-differencing using the results of our Fiducial, High$\A_s$ ($\delta\A_s = +0.05$) and Low$\A_s$ ($\delta\A_s = -0.05$) simulations as
\bq\label{eq:bphi_method_2}
\bphi = \frac{\bphi^{{\rm High}\A_s} + \bphi^{{\rm Low}\A_s}}{2},
\eq
with
\bq\label{eq:bphi_method_3}
\bphi^{{\rm High}\A_s} &=& \frac{4}{+0.05}\Big[\frac{N_g^{\rm High \mathcal{A}_s}}{N_g^{\rm Fiducial}} - 1\Big], \nonumber \\
\bphi^{{\rm Low}\A_s} &=& \frac{4}{-0.05}\Big[\frac{N_g^{\rm Low \mathcal{A}_s}}{N_g^{\rm Fiducial}} - 1\Big],
\eq
and where $N_g$ represents the number of objects in some selection variable bin (total mass, stellar mass, black hole mass, etc.) and the superscripts indicate in which cosmology the number of galaxies is counted.\footnote{In reality, the $N_g$ correspond to the number of galaxies in cosmologies with local PNG, i.e.~the response to $\delta\A_s$ should be evaluated using simulations with local PNG initial conditions, whereas we do so using Gaussian distributed initial conditions. We note, however, that for the currently allowed values of $\fnl$, as well as for the mass scales we consider in this paper, the impact of this approximation is negligible \cite{2000ApJ...541...10M, 2008JCAP...04..014L, 2011JCAP...08..003L}.} The choice of $|\delta\A_s| = 0.05$ is motivated by the compromise between having a sizeable and measurable impact of the change in $\A_s$ on the galaxy abundance, while keeping negligible higher-order corrections to the first-order finite-difference result. Further, having just a single realization of the initial conditions for each cosmology/resolution, it is not possible to estimate our measurement errors in a statistical ensemble sense. As a compromise, we take the difference between the values of $\bphi^{{\rm High}\A_s}$ and $\bphi^{{\rm Low}\A_s}$ (which should be the same up to numerical noise) as our estimate of the error on $\bphi$; see Ref.~\cite{2020JCAP...12..013B} for a discussion of why this yields trustworthy error estimates.

Using the same simulations and methodology, Refs.~\cite{2020JCAP...12..013B, 2020arXiv201204637V} presented an indepth study of the total- and stellar-mass dependence of $\bphi$. Here, we will reproduce some of these past results (while showing also additional ones for other galaxy selection criteria) to better compare with the results for the second-order $\bphidelta$ parameter.

\subsection{The second-order local PNG bias parameter $b_{\phi\delta}$}
\label{sec:method_bphidelta}

The same separate universe simulations can be used to estimate the bias parameter $\bphidelta$ via the response of $P_{gm}(k)$ to long-wavelength primordial potential perturbations $\fnl\phi$. Concretely, in all our cosmologies, the large-scale galaxy-matter cross-power spectrum is described by $P_{gm} = b_1P_{mm}$. Defining its linear local PNG response function as $R_{\phi, gm} = \partial {\rm ln} P_{gm}/\partial(\fnl\phi)$, then it follows that (the power spectrum response functions can be defined in analogy to the galaxy bias parameters by treating the local power spectrum as a biased tracer \cite{responses1})
\bq\label{eq:Rphigm}
R_{\phi, gm} = \frac{\partial{\rm ln}b_1}{\partial(\fnl\phi)} + R_{\phi, mm},
\eq
where $R_{\phi, mm} = \partial {\rm ln} P_{mm}/\partial(\fnl\phi)$. The first term on the right can be worked out by using that
\bq\label{eq:bphidelta}
b_1 = \frac{1}{\bar{n}_g}\frac{\partial \bar{n}_g}{\partial\delta_m}\ \ \ \ ; \ \ \ \bphidelta = \frac{1}{\bar{n}_g}\frac{\partial^2 \bar{n}_g}{\partial(\fnl\phi)\partial\delta_m},
\eq
to yield
\bq\label{eq:Rphigm_2}
R_{\phi, gm} = \frac{\bphidelta}{b_1} - \bphi + R_{\phi, mm},
\eq
where we have used also Eq.~(\ref{eq:bphi_method}). The power spectrum response functions $R_{\phi, gm}$ and $R_{\phi, mm}$ can be measured straightforwardly using the separate universe simulations analogously to how $\bphi$ is estimated in Eqs.~(\ref{eq:bphi_method_2}) and (\ref{eq:bphi_method_3}). For example, using the Fiducial and High$\A_s$ cosmologies, we would estimate the response as $R_{\phi, gm}^{\rm High\A_s} = 4 \left[P_{gm}^{{\rm High}\A_s}/P_{gm}^{\rm Fiducial} - 1\right]/\delta\A_s$. Thus, given the estimates of $b_1$ and $\bphi$ described in the last subsection, as well as measurements of the power spectrum responses $R_{\phi, gm}$ and $R_{\phi, mm}$ on large scales, we can use Eq.~(\ref{eq:Rphigm_2}) to fit for $\bphidelta$. The linear power spectrum response functions effectively describe the mode-coupling structure of squeezed-limit bispectra \cite{2014JCAP...05..048C, response, responses1}, which are sensitive to second-order bias parameters to leading order, and is what allows us to fit for $\bphidelta$ (as well as other second-order bias parameters; cf.~App.~\ref{app:b2}). The parameter $\bphidelta$ can also be estimated from the response of the galaxy power spectrum $P_{gg} = b_1^2 P_{mm} + 1/\bar{n}_g$, although in this case the contribution from shot noise lowers the signal-to-noise unnecessarily; further, it can be straightforwardly shown that the response of this galaxy power spectrum model to $\fnl\phi$ agrees with the expressions shown in Ref.~\cite{2020JCAP...10..007C} obtained with squeezed-limit bispectra.

The results on $\bphidelta$ that we will show below were obtained, however, with a different (yet related) strategy. Rather than using the separate universe simulations to differentiate $P_{gm}$ and $P_{mm}$, we first fit instead for $b_1$ in the three cosmologies separately using Eq.~(\ref{eq:b1_method}), and then we differentiate $b_1$ using finite differences. With this estimate, together with the estimates of $b_1$ and $\bphi$, $\bphidelta$ is then simply given by
\bq\label{eq:bphidelta_2}
\bphidelta = \bigg[\frac{\partial{\rm ln}b_1}{\partial(\fnl\phi)} + \bphi \bigg] b_1.
\eq
We have explicitly checked that both strategies above give consistent results, but the latter yielded slightly better signal-to-noise estimates, which is why we adopt it as the default. Concretely, we evaluate the response of $b_1$ as 
\bq\label{eq:respb1}
\frac{\partial{\rm ln}b_1}{\partial(\fnl\phi)} = \frac{1}{2}\left[\frac{\partial{\rm ln}b_1}{\partial(\fnl\phi)}\right]^{{\rm High}\A_s} + \frac{1}{2}\left[\frac{\partial{\rm ln}b_1}{\partial(\fnl\phi)}\right]^{{\rm Low}\A_s},
\eq
with
\bq\label{eq:respb1_2}
\left[\frac{\partial{\rm ln}b_1}{\partial(\fnl\phi)}\right]^{{\rm High}\A_s} &=& \frac{4}{+0.05}\Big[\frac{b_1^{\rm High \mathcal{A}_s}}{b_1^{\rm Fiducial}} - 1\Big], \nonumber \\
\left[\frac{\partial{\rm ln}b_1}{\partial(\fnl\phi)}\right]^{{\rm Low}\A_s} &=& \frac{4}{-0.05}\Big[\frac{b_1^{\rm Low \mathcal{A}_s}}{b_1^{\rm Fiducial}} - 1\Big],
\eq
and we estimate its error analogously to as for $\bphi$. The final uncertainty on $\bphidelta$ is worked out by standard propagation of uncertainty in Eq.~(\ref{eq:bphidelta_2}) assuming uncorrelated errors, which is conservative since $\bphi$, $b_1$ and its response are measured from the same simulations and so their uncertainties are correlated. In App.~\ref{app:b2} we validate this strategy to estimate $\bphidelta$ by applying it (with the appropriate modifications described there) to estimate the second-order bias parameter $b_2$,  whose values can be compared with known results in the literature. 

We note before proceeding that the $\bphidelta$ parameter can also be estimated by fitting perturbation theory models to the bispectrum measured from simulations with local PNG initial conditions \cite{2021JCAP...05..015M}. This requires however very large simulation volumes (in fact still currently out of reach for self-consistent galaxy formation simulations) in order to measure the bispectrum precisely on large-scales where the effects of $\fnl$ dominate. At fixed volume, separate universe simulations offer thus the ideal method to study the local PNG bias parameters. Within the separate universe approach, there is an alternative way to estimate $\bphidelta$ using separate universe simulations that incorporate simultaneously the effects of total mass and primordial gravitational potential perturbations. This would require however additional simulations for higher amplitudes of the long-wavelength modes to be sensitive to the second-order terms in the galaxy bias expansion, similarly to how Ref.~\cite{lazeyras/etal} uses separate universe simulations to estimate higher-order bias parameters such as $b_2$ and $b_3$.

\section{Galaxy bias results}
\label{sec:results}

\begin{figure}
\centering
\includegraphics[width=\textwidth]{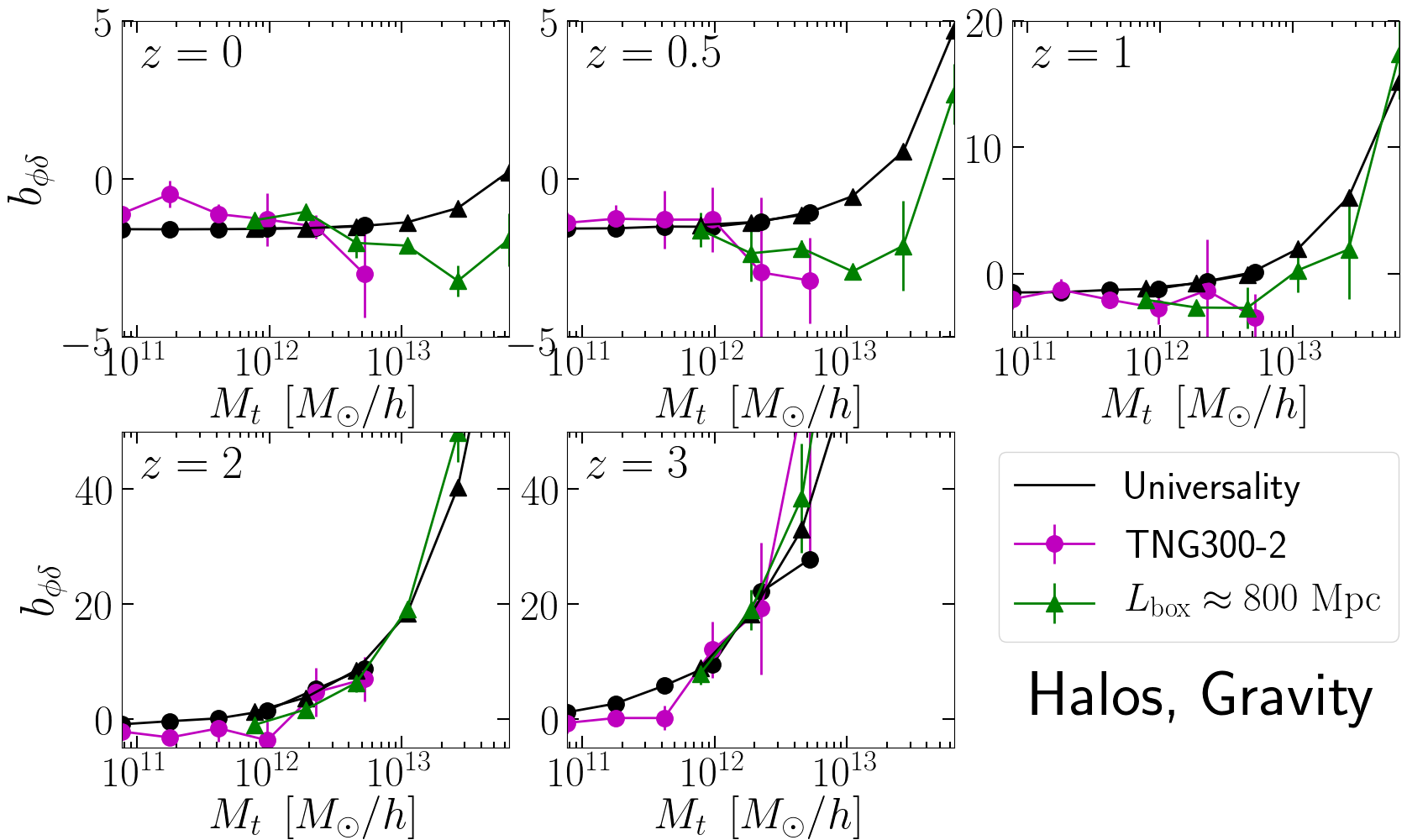}
\caption{The total-mass dependence of $\bphidelta$ for dark matter halos in the gravity-only simulations. The different panels are for different redshifts and the colored points with error bars show the results from the TNG300-2 and $\bigbox$ simulations, as labeled. The result in black shows the prediction of the universality relation of Eq.~(\ref{eq:bphidelta_uni}), $\bphidelta = b_\phi - b_1 + 1 + \delta_c[b_2 - (8/21)(b_1-1)]$, using the values of $b_1$ measured for the same halos, with $\bphi = 2\delta_c(b_1-1)$ and $b_2(b_1)$ given by Eq.~(\ref{eq:b2titouan}).}
\label{fig:bphidelta_totmass}
\end{figure}

In this section we show and discuss our main numerical results on the galaxy bias parameters $\bphi$ and $\bphidelta$. We begin with the $\bphidelta$ parameter measured for halos and subhalos in the gravity-only simulations, and then compare the results of the $\bphi(b_1)$ and $\bphidelta(b_1)$ relations for a number of different galaxy samples in the IllustrisTNG simulations; we show results for galaxies selected by total mass $M_t$, stellar mass $M_*$, black hole mass $M_{\rm BH}$, black hole mass accretion rate $\dot{M}_{\rm BH}$ and $(g-r)$ color.

\subsection{The $b_{\phi\delta}$ parameter in gravity-only simulations}
\label{sec:results_gravity}

The total-mass dependence of $\bphidelta$ is shown in Fig.~\ref{fig:bphidelta_totmass} for dark matter halos at different redshifts, and for the gravity-only TNG300-2 and $\bigbox$ simulations, as labeled. The result is compared to the universality prediction of Eq.~(\ref{eq:bphidelta_uni}) (shown in black), evaluated using the values of $b_1$ estimated for the same mass bins, together with the universality relation for the linear local PNG parameter $\bphi = 2\delta_c(b_1-1)$ and $b_2(b_1)$ given by the fit obtained for halos with separate universe simulations in Ref.~\cite{lazeyras/etal}:
\bq\label{eq:b2titouan}
b_2(b_1) = 0.412 - 2.143b_1 + 0.929b_1^2 + 0.008b_1^3.
\eq
The figure shows that the $\bphidelta$ values measured in the simulations depart from the universality expectation, with the difference being both redshift- and mass-dependent. Concretely, for $M_t \lesssim 10^{12}M_{\odot}/h$, there is a slight trend for $\bphidelta$ to overpredict (less negative) the universality prediction at $z\leq0.5$, but to underpredict it at $z \geq 2$ (more negative). On the other hand, for $M_t \gtrsim 10^{12}M_{\odot}/h$, the universality relation prediction is visibly above the measured $\bphidelta$ at $z\leq1$, but the two are consistent within the errors at $z \geq 2$. As one would expect for its smaller volume, the TNG300-2 results appear noisier compared to $\bigbox$, but it is nonetheless possible to discern a good overall agreement between the two resolutions on the mass scales where they overlap.

\begin{figure}
\centering
\includegraphics[width=\textwidth]{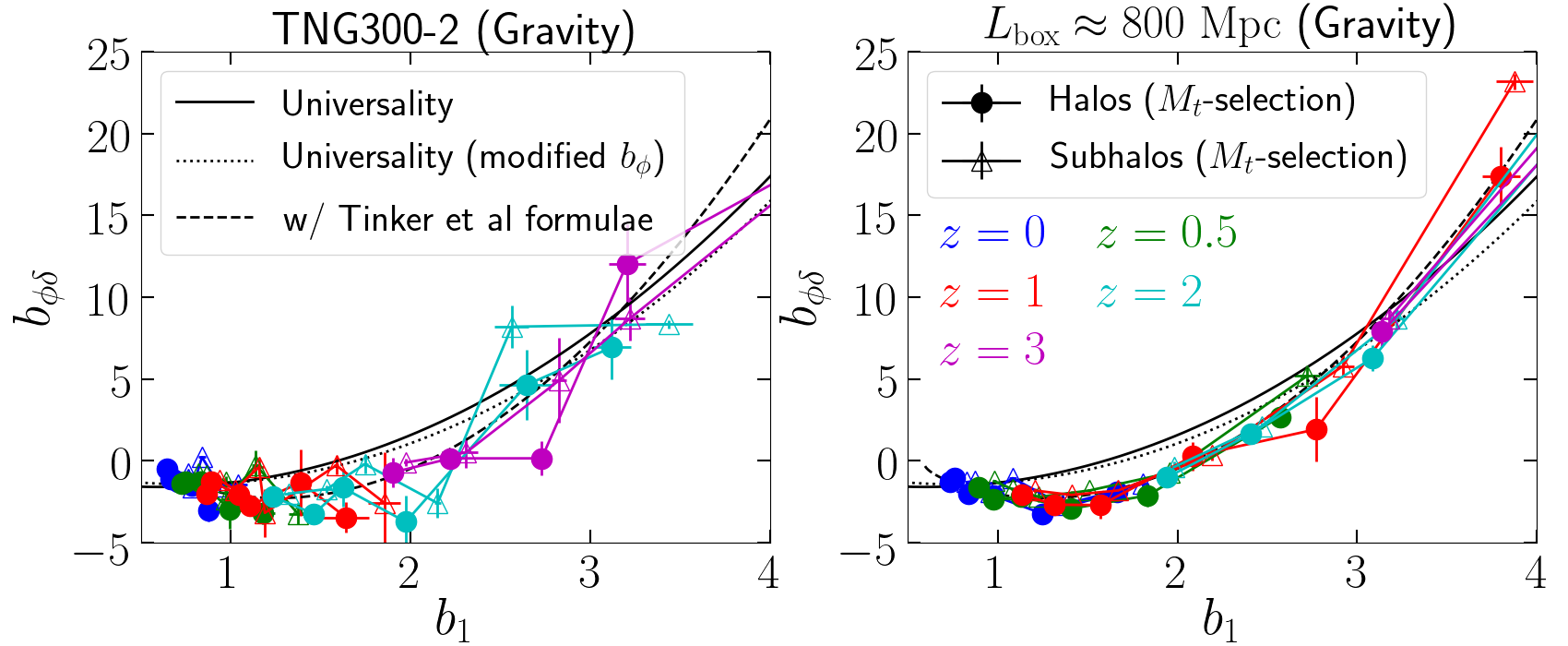}
\caption{The $\bphidelta(b_1)$ relation for halos (filled circles) and subhalos (open triangles) in the gravity-only simulations. Each data point shows $\bphidelta$ for the mass bins in Fig.~\ref{fig:bphidelta_totmass}, but now plotted against $b_1$ in the same mass bin. The two panels are for the TNG300-2 and $\bigbox$ simulations and the different colors show the result at different redshifts, as labeled. In both panels, the solid black line shows the universality prediction of Eq.~(\ref{eq:bphidelta_uni}), $\bphidelta = b_\phi - b_1 + 1 + \delta_c[b_2 - (8/21)(b_1-1)]$, with $\bphi = 2\delta_c(b_1-1)$ and $b_2(b_1)$ given by Eq.~(\ref{eq:b2titouan}). The dotted line shows the same but with $\bphi = 0.85 \times 2\delta_c(b_1-1)$ to account for the departures from universality on this relation. {The dashed line shows the outcome of Eq.~(\ref{eq:bphidelta_2}) obtained using the formulae from Tinker et al \cite{2008ApJ...688..709T, 2010ApJ...724..878T}.}}
\label{fig:bphidelta_b1}
\end{figure}

Figure \ref{fig:bphidelta_b1} shows the $\bphidelta(b_1)$ relation for total-mass selected halos and subhalos in the gravity-only TNG300-2 and $\bigbox$ simulations at different redshifts, as labeled. The relation is seen to depend only weakly on redshift; this can be better appreciated in the higher signal-to-noise results from the $\bigbox$ box, but the TNG300-2 results display a consistent picture. Further, in the range $1 \lesssim b_1 \lesssim 3$, the $\bphidelta$ values of the simulations are systematically below (more negative/less positive) the universality prediction shown by the solid black line; the largest difference occurs at $b_1 \approx 2$ and is $\Delta \bphidelta \sim 3$. For $b_1 \lesssim 0.8$, one can also discern a trend for the measured $\bphidelta$ to overpredict the universality relation, which is a manifestation of the same trend shown in Fig.~\ref{fig:bphidelta_totmass} at low redshift and lower masses. Note also that the results for halos and subhalos are consistent within the precision of our measurements, although some differences are in general to be expected (even if small) since the relation between $\bphidelta(b_1)$ is nonlinear (see Sec.~3.3 of Ref.~\cite{2021arXiv210502876B} for a discussion).

The dotted lines in Fig.~\ref{fig:bphidelta_b1} show the prediction of a variant of the universality relation of Eq.~(\ref{eq:bphidelta_uni}), in which instead of using the universality relation for $\bphi$ (cf.~Eq.~(\ref{eq:bphi_uni})), one replaces it by $\bphi(b_1) = 0.85 \times 2\delta_c(b_1-1)$, which offers a more adequate approximation to the $\bphi(b_1)$ relation in gravity-only simulations \cite{grossi/etal:2009, desjacques/seljak/iliev:2009, 2010MNRAS.402..191P, 2011PhRvD..84h3509H, 2017MNRAS.468.3277B, 2020JCAP...12..013B} (see the dotted line in the top left panel of Fig.~\ref{fig:bphi_bphidelta_all} below). This variant of the universality relation with a modified $\bphi(b_1)$ relation does get slightly closer to the simulation measurements over the range $1 \lesssim b_1 \lesssim 3$, but not sufficiently to bring the two results into agreement. In other words, the breakdown of the universality relation for the second-order parameter $\bphidelta$ cannot be attributed solely to the breakdown of the universality relation for the linear parameter $\bphi$. {The dashed lines in Fig.~\ref{fig:bphidelta_b1} show the outcome of Eq.~(\ref{eq:bphidelta_2}) evaluated with the halo mass function and halo bias $b_1$ formulae from Tinker et al \cite{2008ApJ...688..709T, 2010ApJ...724..878T}. Concretely, the $\partial{\rm ln}b_1/\partial(\fnl\phi) \equiv 4\partial{\rm ln}b_1/\partial(\delta\A_s)$ term is evaluated by finite-differencing the $b_1$ formula w.r.t.~$\A_s$, and $\bphi$ is obtained by finite-differencing the halo mass function also w.r.t~$\A_s$ as in Eq.~(\ref{eq:bphi_method}) (this aproach to $\bphi$ compares well with $\bphi = 0.85 \times 2\delta_c(b_1-1)$). This semi-analytical calculation agrees extremelly well with our numerical estimates, which represents a good cross-check of both results. Finally, we have also found the following best-fitting quadratic polynomial to the dark matter halo results (fitted up to $b_1<4$ using all redshifts and both simulation boxes):
\bq\label{eq:quadfit}
\bphidelta(b_1)^{\rm fit} = 3.85 - 9.49b_1 + 3.44b_1^2,
\eq
but which we skip showing to avoid crowing the figure.}

\subsection{The $b_{\phi}$ and $b_{\phi\delta}$ parameters in galaxy formation simulations}
\label{sec:results_hydro}

\begin{figure}
\centering
\includegraphics[width=\textwidth]{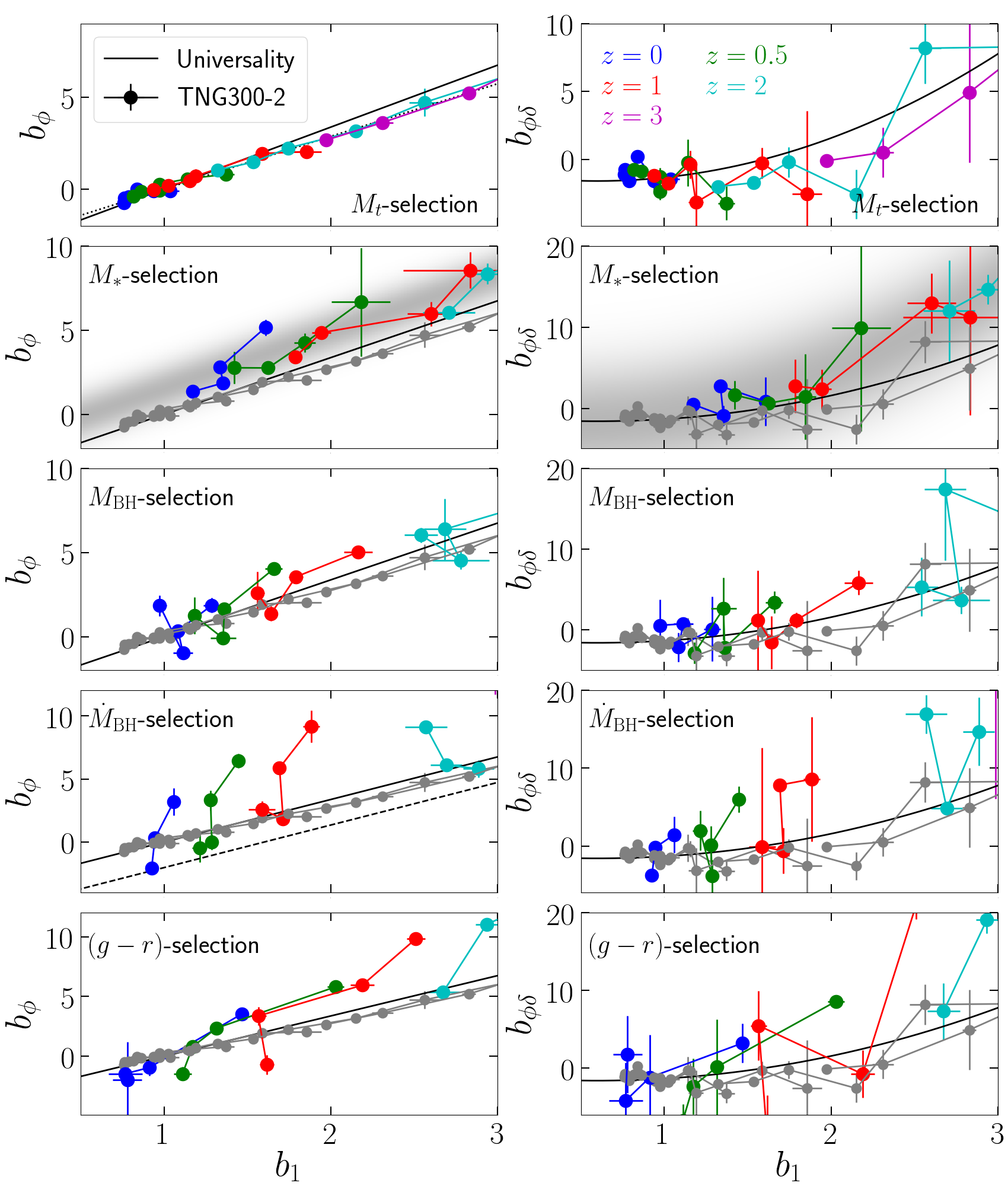}
\caption{The $\bphi(b_1)$ and $\bphidelta(b_1)$ relations for IllustrisTNG galaxies. The left and right panels are for $\bphi$ and $\bphidelta$, the colors indicate the redshift, and from top to bottom, the panels show the result for galaxies selected by their total mass, stellar mass, black hole mass, black hole mass accretion rate and dust-uncorrected $(g-r)$ color, as labeled (the $M_t$ results for $\bphidelta$ are shown also in Fig.~\ref{fig:bphidelta_b1}, but are repeated here to ease comparisons). Each data point shows the value of $\bphi$ and $\bphidelta$ for the galaxies in some property bin, plotted against the value of $b_1$ for the same galaxies. The solid lines show the corresponding universality predictions (cf.~Eqs.~(\ref{eq:bphi_uni}) and (\ref{eq:bphidelta_uni})), and to ease comparisons, the grey points repeat the result of the top panels for total-mass selection. The $\bphi(b_1)$ panels for $M_t-$ and $\dot{M}_{\rm BH}$-selection show also the variants of the universality relation $\bphi(b_1) = 0.85 \times 2\delta_c(b_1 - 1)$ (dotted) and $\bphi(b_1) = 2\delta_c(b_1 - 1.6)$ (dashed), respectively. The grey color maps in the $M_*$ panels indicate the amplitude of the Gaussian priors of Eqs.~(\ref{eq:prior_bphi}) and (\ref{eq:prior_bphidelta}). This figure displays results only on the observationally interesting range $1 \lesssim b_1 \lesssim 3$, but see Fig.~\ref{fig:vsselections} for bias parameter values that do not appear here.}
\label{fig:bphi_bphidelta_all}
\end{figure}

\begin{figure}
\begin{subfigure}
\centering
\includegraphics[width=\textwidth]{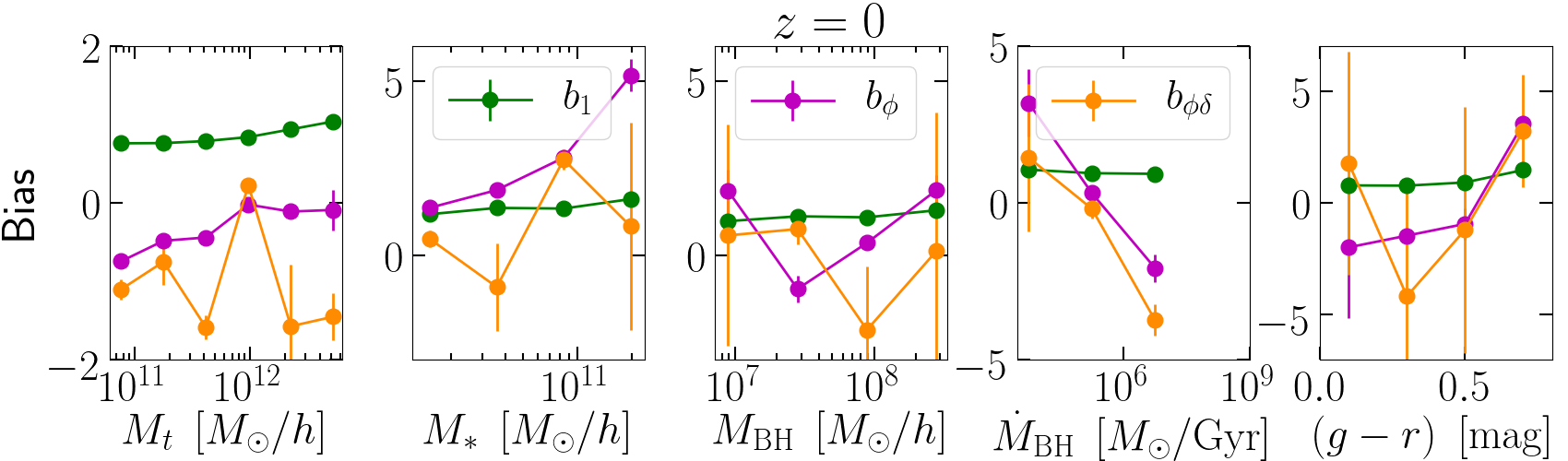}
\end{subfigure}
\begin{subfigure}
\centering
\includegraphics[width=\textwidth]{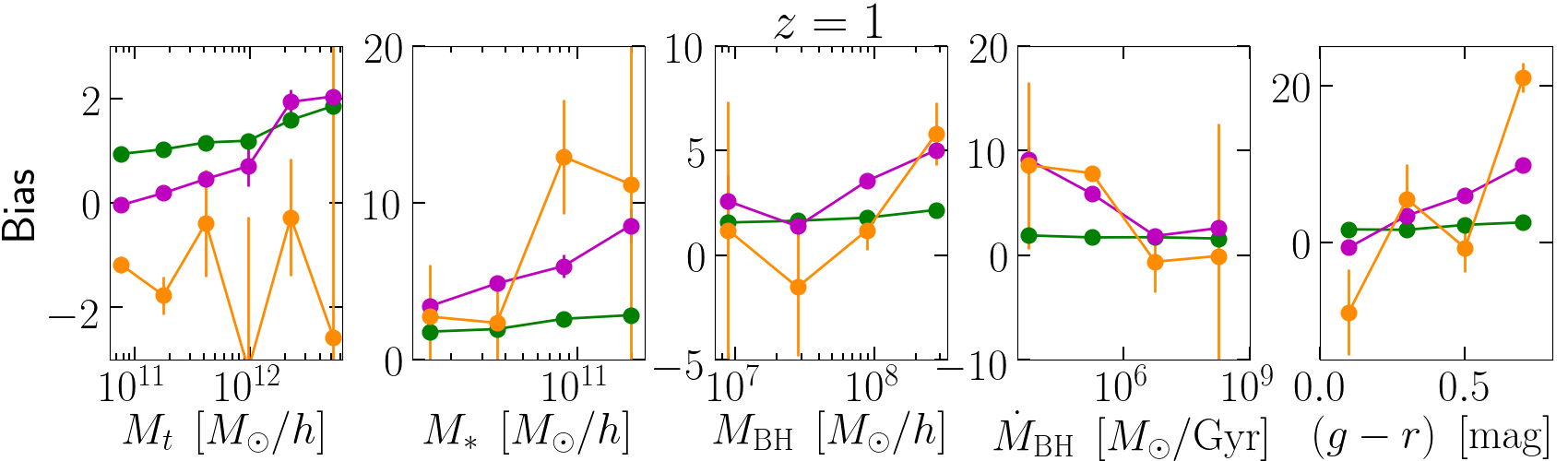}
\end{subfigure}
\begin{subfigure}
\centering
\includegraphics[width=\textwidth]{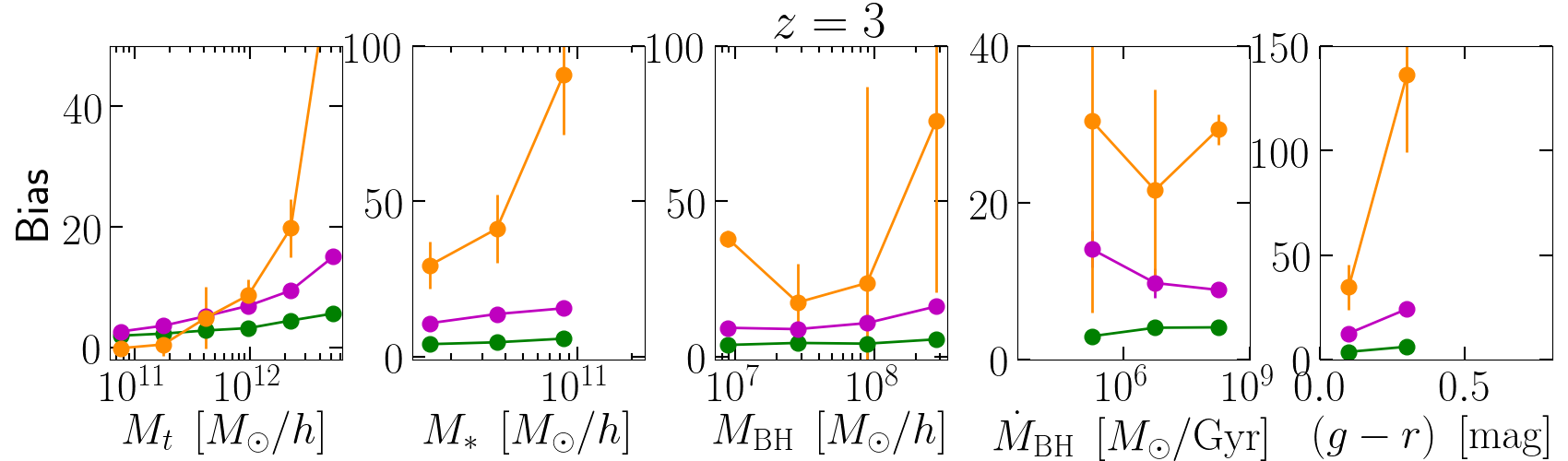}
\end{subfigure}
\caption{The dependence of the bias parameters $b_1$ (green), $\bphi$ (magenta) and $\bphidelta$ (orange) on the galaxy total mass $M_t$, stellar mass $M_*$, black hole mass $M_{\rm BH}$, black hole mass accretion rate $\dot{M}_{\rm BH}$ and dust-uncorrected $(g-r)$ color, as labeled. From top to bottom, the panels show the result for $z=0$, $z=1$ and $z=3$. This figure helps understand some of the trends seen in the main result of Fig.~\ref{fig:bphi_bphidelta_all} on the $\bphi(b_1)$ and $\bphidelta(b_1)$ relations.}
\label{fig:vsselections}
\end{figure}

We turn our attention now to the $\bphi$ and $\bphidelta$ parameters measured from the hydrodynamical, full-physics IllustrisTNG runs with the TNG300-2 box; in this subsection we focus only on the results for galaxies (i.e., subhalos with mass in stars). Figure \ref{fig:bphi_bphidelta_all} shows the $\bphi(b_1)$ and $\bphidelta(b_1)$ relations at different redshifts for galaxies selected by their total mass $M_t$, stellar mass $M_*$, black hole mass $M_{\rm BH}$, black hole mass accretion rate $\dot{M}_{\rm BH}$, and dust-uncorrected $(g-r)$ color, as labeled. Concretely, we consider $6$ $M_t$ bins log-spaced between $\left[5\times 10^{10}; 8 \times 10^{12}\right]M_{\odot}/h$. For $M_*$, $M_{\rm BH}$ and $\dot{M}_{\rm BH}$ we consider 4 bins log-spaced between $\left[1\times 10^{10}; 3 \times 10^{11}\right]M_{\odot}/h$, $\left[5\times 10^{6}; 5 \times 10^{8}\right]M_{\odot}/h$ and $\left[1\times 10^{3}; 1 \times 10^{9}\right]M_{\odot}/{\rm Gyr}$, respectively. And for $(g-r)$ we consider 4 bins linearly spaced within $\left[0, 0.8\right]$. In Fig.~\ref{fig:vsselections}, we show the dependence of the three galaxy bias parameters $b_1$, $\bphi$ and $\bphidelta$ on the galaxy selection variables themselves for $z=0$, $z=1$ and $z=3$.

For the case of $\bphi$, the total-mass and stellar-mass results in Fig.~\ref{fig:bphi_bphidelta_all} (top two panels on the left) have been already discussed in detail in Ref.~\cite{2020JCAP...12..013B} using the same simulations as in this paper. Namely, for $M_t$-selection, the figure shows the well-known result that the $\bphi(b_1)$ relation of the simulated objects is below the universality expectation for $b_1 \gtrsim 1.5$, and that the modified relation $\bphi(b_1) = 0.85 \times 2\delta_c(b_1 - 1)$ (shown by the dotted line) provides a more adequate description. For $M_*$-selection, the $\bphi(b_1)$ relation of the galaxies systematically overpredicts the universality relation, with $\bphi(b_1) = 2 \delta_c(b_1 - 0.55)$ being a more faithful approximation of the simulation measurements (center of the grey color map). The lower three panels on the left of Fig.~\ref{fig:bphi_bphidelta_all} display the result for galaxies selected by $M_{\rm BH}$, $\dot{M }_{\rm BH}$ and $(g-r)$, and they illustrate the strong sensitivity of the $\bphi(b_1)$ relation to the galaxy selection criterion adopted, as well as the fact that the universality relation continues to be an inadequate description. We highlight for example the case of the $\dot{M}_{\rm BH}$-selected galaxies, for which the $\bphi(b_1)$ relation becomes especially steep, and does not even admit a regular, redshift-independent function of $b_1$.

The dashed line in the $\dot{M }_{\rm BH}$ panel for $\bphi$ shows the prediction of the variant of the universality relation $\bphi(b_1) = 2\delta_c(b_1 - 1.6)$, which was derived by Ref.~\cite{slosar/etal:2008} for dark matter halos that had undergone a recent major merger. The authors further hypothesized it could be a better description of the $\bphi(b_1)$ relation of quasars compared to the universality relation, as recent mergers may correlate with strong active galactic nuclei (AGN) activity/luminosity. The reason why it is interesting to make this comparison is because AGN luminosity is also thought to be proportional to the accretion rate of the supermassive black hole. The result of Fig.~\ref{fig:bphi_bphidelta_all} shows, however, that at least in the IllustrisTNG model, the relation $\bphi(b_1) = 2\delta_c(b_1 - 1.6)$ is not a good description of the objects selected by $\dot{M }_{\rm BH}$ (or by proxy, selected by their AGN luminosity). With just a single realization of the initial conditions of the simulations, and for a single galaxy formation model, our results do now allow us yet to conclude decisively on the $\bphi(b_1)$ relation of real-life quasars and AGN. However, in order to obtain competitive and unbiased constraints on $\fnl$, the $\dot{M }_{\rm BH}$ results depicted in Fig.~\ref{fig:bphi_bphidelta_all} do strongly motivate more works with galaxy formation separate universe simulations to determine the precise bias relations for these objects. Note that the tightest constraints on $\fnl$ using large-scale structure to date were obtained precisely with quasar samples from eBOSS DR14 \cite{2019JCAP...09..010C} and DR16 \cite{2021arXiv210613725M}, who assumed $\bphi(b_1) = 2\delta_c(b_1 - 1.6)$ in parts of their analysis. 

For the case of the $\bphidelta$ results on the right of Fig.~\ref{fig:bphi_bphidelta_all}, a first point to note concerns the lower signal-to-noise of the measurements compared to $\bphi$. This is not surprising since $\bphidelta$ is a second-order bias parameter, and our method to estimate it relies on estimating $b_1$ using the large-scale limit of the galaxy-matter cross-power spectrum (cf.~Sec.~\ref{sec:method_bphidelta}), which can be somewhat uncertain on a $L_{\rm box} = 205{\rm Mpc}/h$ box; recall from Fig.~\ref{fig:bphidelta_b1} how the signal-to-noise improves substantially from the TNG300-2 to the bigger $\bigbox$ box. There are nonetheless a few trends that one can discern. For example, similarly to the $\bphi(b_1)$ relation, there is also a visible trend for the $\bphidelta$ values at fixed $b_1$ to be higher for stellar-mass relative to total-mass selection, and the $\bphidelta(b_1)$ relation displays similar {\it steep} variations for $\dot{M}_{\rm BH}$-selected galaxies, i.e., $\bphidelta$ can vary rapidly in a narrow $b_1$-interval. Also similarly to $\bphi$, the $\bphidelta(b_1)$ relations do not show evidence of being robust to changes in the galaxy selection criterion, although the poorer signal-to-noise here makes it more difficult to draw decisive conclusions.

The details of the $\bphi(b_1)$ and $\bphidelta(b_1)$ relations can be understood by inspecting the corresponding dependencies of the bias parameters on the galaxy selection variables in Fig.~\ref{fig:vsselections}. For example, the shape of the dependence of the bias parameters can vary quite significantly  from one galaxy property to another, which is the reason behind the strong sensitivity of the $\bphi(b_1)$ and $\bphidelta(b_1)$ relations to the galaxy selection criteria seen in Fig.~\ref{fig:bphi_bphidelta_all}. It is interesting to contrast this result for $\bphi(b_1)$ and $\bphidelta(b_1)$, with the appreciably weaker sensitivity shown in Ref.~\cite{2021arXiv210502876B} for the $b_2(b_1)$ and $b_{K^2}(b_1)$ relations using also IllustrisTNG galaxies.  As described in Refs.~\cite{2020JCAP...12..013B, 2020arXiv201204637V, 2021arXiv210502876B}, this can be explained using the halo model and halo occupation distribution formalisms, but we leave a more detailed investigation along these lines to future work.

Finally, the grey color maps in the stellar-mass panels in Fig.~\ref{fig:bphi_bphidelta_all} describe the shape of Gaussian priors on the $\bphi(b_1)$ and $\bphidelta(b_1)$ relations that we will use in the next section to study the impact of galaxy bias uncertainties on $\fnl$ constraints. Concretely, for $\bphi$ and $\bphidelta$, our assumed priors are, respectively, 
\bq
\label{eq:prior_bphi}           \P(\bphi | b_1) &=& {\rm exp}\left[-\frac{1}{2}\frac{\left(\bphi - \mu_{\bphi}(b_1)\right)^2}{\Delta{\bphi}^2}\right] \ \ \ ;\ \mu_{\bphi}(b_1) = 2\delta_c\left(b_1-0.55\right)\ \ \ \ \ \ \ \ \ \ ;\ \Delta{\bphi} = 1; \\
\label{eq:prior_bphidelta} \P(\bphidelta | b_1) &=& {\rm exp}\left[-\frac{1}{2}\frac{\left(\bphidelta - \mu_{\bphidelta}(b_1)\right)^2}{\Delta{\bphidelta}^2}\right] \ ;\ \mu_{\bphidelta}(b_1) =  -1.7 - 1.6b_1 + 2.4b_1^2\ \ ;\ \Delta{\bphidelta} = 5;
\eq
where, we stress, the mean relations are expected to describe only the results for $M_*$-selection, and the standard deviations are assumed $b_1$-independent for simplicity and chosen to roughly match the uncertainty in our numerical results with IllustrisTNG.

\section{Impact of galaxy bias uncertainties on $\fnl$ constraints}
\label{sec:forecasts}

In this section we show and discuss the results of an idealized forecast setup for a fictitious survey, to study the impact that uncertainties on $\bphi$ and $\bphidelta$ have on $\fnl$ constraints. This analysis effectively extends that of Ref.~\cite{2020JCAP...12..031B} to include also uncertainties on the $\bphidelta(b_1)$ relation.

\subsection{Forecast setup}
\label{sec:forecasts_setup}

We work with a Gaussian likelihood function with a data vector consisting of hypothetical measurements of the real-space multitracer power spectrum $P_{gg}$ \cite{2009JCAP...10..007M, 2009PhRvL.102b1302S} and bispectrum $B_{ggg}$
\bq\label{eq:datavector}
\vD = \Big\{ \hat{P}_{gg}^{\rm AA} , \hat{P}_{gg}^{\rm AB} , \hat{P}_{gg}^{\rm BB} , \hat{B}_{ggg}^{\rm AAA}\Big\},
\eq
where the superscripts A and B indicate two galaxy samples, e.g., $\hat{P}_{gg}^{\rm AB}$ denotes the cross-power spectrum of the two samples. {We consider only the bispectrum of sample A for simplicity; a more involved analysis could include also the bispectrum of sample B and the cross-bispectra $B_{ggg}^{\rm AAB}$ and $B_{ggg}^{\rm ABB}$, but this additional complication is not critical to our discussion.}

In our theory model, we evaluate the galaxy power spectrum and bispectrum at tree level using the following galaxy bias expansion
\bq\label{eq:biasexp}
\delta_g(\vx, z) &=& b_1(z)\delta_m(\vx, z) + \frac{1}{2} b_2(z) [\delta_m(\vx, z)]^2 + b_{K^2}(z)[K_{ij}(\vx, z)]^2 + \eps(\vx) + \eps_\delta(\vx)\delta_m(\vx, z) \nonumber \\
&+& \fnl \big[b_\phi(z) \phi(\vq) + b_{\phi\delta}(z)\phi(\vq)\delta_m(\vx, z) + \eps_\phi(\vx)\phi(\vq)\big],
\eq
where in addition to the deterministic terms that appeared already before, we include now also the relevant stochastic contributions $\eps, \eps_\delta, \eps_\phi$. This expansion contains all terms that are needed to self-consistently derive the leading-order galaxy power spectrum and bispectrum \cite{assassi/baumann/schmidt}. We display the corresponding expressions for the power spectrum, bispectrum and our treatment of the covariance matrix in App.~\ref{app:theory}.

We consider a galaxy sample at redshift $z=1$ covering a volume of $V = 100{\rm Gpc^3}/h^3$. For the multitracer part of the data vector, we split this galaxy sample into a low- and a high-stellar-mass subsamples, with $M_* \in \left[5\times10^{10} ; 2\times10^{11}\right]M_{\odot}/h$ for subsample A and $M_* > 2\times10^{11}M_{\odot}/h$ for subsample B. For the stellar mass function of the IllustrisTNG simulations this corresponds to the number densities $\bar{n}_g^{\rm A} = 1.74\times10^{-3}\left[{h^3/{\rm Mpc^3}}\right]$, $\bar{n}_g^{\rm B} = 1.07\times10^{-4}\left[{h^3/{\rm Mpc^3}}\right]$, and linear bias parameters $b_1^{\rm A} = 1.58$, $b_1^{\rm B} = 2.37$. We evaluate $b_2$ and $b_{K^2}$ for these samples using $b_2(b_1) = 0.30 - 0.79b_1 + 0.20b_1^2 + 0.12b_1^3$ and $b_{K^2}(b_1) = 0.66 - 0.57b_1$, which are fits for IllustrisTNG galaxies obtained by Ref.~\cite{2021arXiv210502876B}. The fiducial values of $\bphi$ and $\bphidelta$ are given by the mean relations in Eqs.~(\ref{eq:prior_bphi}) and (\ref{eq:prior_bphidelta}). We assume Poissonian statistics for the fiducial power spectrum and bispectrum of the noise terms, i.e., $P_{\eps\eps} = 1/\bar{n}_g$, $P_{\eps\eps_\delta} = b_1/(2\bar{n}_g)$ and $B_{\eps\eps\eps} = 1/\bar{n}_g^2$ (but note that we sample these in our constraints too; see Ref.~\cite{2021JCAP...05..015M} for the importance of $k$-dependent corrections to the shot noise in $\fnl$ constraints).

We generate our data vector as a noiseless realization of our theory model for the bias parameters and noise terms listed above, and at a fiducial cosmology that has the same parameters as our Fiducial simulations (cf.~Sec.~\ref{sec:method_sepuni}), except for $\fnl = 5$. The adoption of a noiseless data vector is naturally idealized, but which we note is the most adequate choice for our purpose here to isolate the impact of galaxy bias uncertainties; see also Ref.~\cite{2021JCAP...05..015M}, who takes halo power spectra and bispectra measurements from simulations as the data vector, and so the assessment of the true impact of bias uncertainties gets complicated by the limitations of the theory model to describe the simulation measurements. We consider 42 $k$-bins between $k_{\rm min} = \pi/V_s^{1/3}$ and $k_{\rm max} = 0.2\ h/{\rm Mpc}$ (in intervals of $k_{\rm min}$ up to $0.01h/{\rm Mpc}$ and of $10\times k_{\rm min}$ beyond that).

Similarly to Ref.~\cite{2020JCAP...12..031B}, we show results for two ways to deal with galaxy bias uncertainties:
\begin{itemize}
\item \underline{Parametrization 1: direct priors on $b_\phi$ and $\bphidelta$.} 

In this case, we sample the following 13 dimensional parameter space:
\bq\label{eq:params1}
\vtheta = \{\fnl, b_1^{\rm A}, b_1^{\rm B}, P_\eps^{\rm A}, P_\eps^{\rm B}, \bphi^{\rm A}, \bphi^{\rm B}, {\mathcal{A}_s}, b_2^{\rm A}, b_{K^2}^{\rm A}, \bphidelta^{\rm A}, P_{\eps\eps_\delta}^{\rm A} , B_{\eps\eps\eps}^{\rm A} \}.
\eq
With this parametrization we will study the impact that different priors on $\bphi(b_1)$ and $\bphidelta(b_1)$ have on $\fnl$ constraints.

\item \underline{Parametrization 2: fit for products $\fnl b_\phi$, $\fnl \bphidelta$.} 

In this case, rather than making assumptions on the bias parameters, we fit directly for the parameter combinations $\fnl b_\phi$ and $\fnl b_{\phi\delta}$ as they enter the galaxy power spectrum and bispectrum (cf.~App.~\ref{app:theory}). This approach makes it harder to constrain $\fnl$ directly, but note it can still be powerful as it can provide evidence for $\fnl \neq 0$ in a way that is independent of galaxy bias assumptions. {This parametrization prevents however direct comparisons of the constraining power of galaxy and CMB data, as well as the combination of galaxy- and CMB-based bounds to obtain a tighter combined bound on $\fnl$.} The parameter space is also 13 dimensional:
\bq\label{eq:params2}
\vtheta = \{[\fnl b_\phi^{\rm A}], [\fnl b_\phi^{\rm B}], b_1^{\rm A}, b_1^{\rm B}, P_\eps^{\rm A}, P_\eps^{\rm B}, \fnl, {\mathcal{A}_s},  b_2^{\rm A}, b_{K^2}^{\rm A}, [\fnl b_{\phi\delta}^{\rm A}], P_{\eps\eps_\delta}^{\rm A} , B_{\eps\eps\eps}^{\rm A} \}.
\eq
In this part of the analysis, we will also consider cases in which we do not fit for $[\fnl b_{\phi\delta}^{\rm A}]$, and replace it with priors on $\bphidelta(b_1)$ to illustrate the breaking of parameter degeneracies.
\end{itemize}
The majority of our results are for the full data vector of Eq.~(\ref{eq:datavector}), but we shall also display results for power-spectrum-only analyses (shown in dashed-black and labeled as $P_{gg}$-only). In this case, in parametrizations 1 and 2 we need to consider only the first 7 and 6 parameters, respectively; we keep ${\mathcal{A}_s}$ fixed at the fiducial value in these cases. Except for the $\bphi$ and $\bphidelta$ parameters, we always assume wide uninformative linear priors when sampling the parameter space, which we do using the {\tt EMCEE} {\tt Python} implementation \cite{2013PASP..125..306F} of the affine-invariant Markov Chain Monte Carlo (MCMC) sampler in Ref.~\cite{2010CAMCS...5...65G}. We use 32 {\it walkers} with a chain convergence criteria that (i) the size of the chain must be $100$ times the autocorrelation time and (ii) the latter having varied less than $1\%$ since the last calculation point, which is every few thousand samples.

{We stress that our idealized forecast setup serves primarily the purpose to provide a simple framework to visualize the impact that galaxy bias uncertainties can have on $\fnl$ constraints, and that it is not intended to be representative of any specific current or future survey.}

\subsection{Results from parametrization 1: direct priors on $b_\phi$ and $\bphidelta$}
\label{sec:forecasts_param1}

\begin{figure}
\centering
\includegraphics[width=\textwidth]{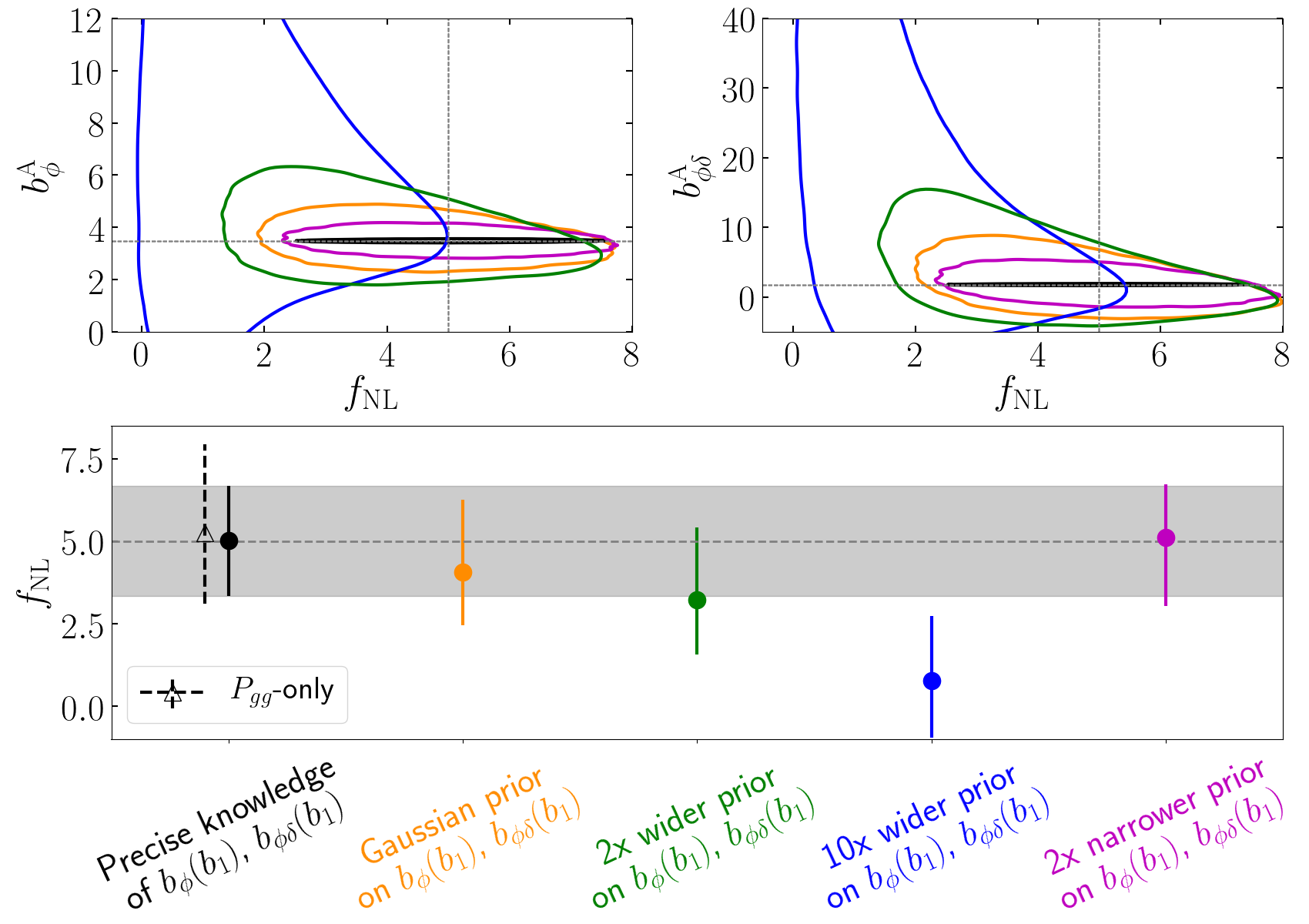}
\caption{Constraint results from parametrization 1 with direct priors on the $\bphi(b_1)$ and $\bphidelta(b_1)$ relations. The upper panels show the two-dimensional marginalized $1\sigma$ constraints on the $\bphi^{\rm A}-\fnl$ and $\bphidelta^{\rm A}-\fnl$ planes, and the lower panel shows the marginalized $1\sigma$ constraints on $\fnl$. The different colors are for different adopted priors on $\bphi(b_1)$ and $\bphidelta(b_1)$: the black lines show the result assuming perfect knowledge of the fiducial $\bphi(b_1)$ and $\bphidelta(b_1)$ relations, the results in orange are for the Gaussian priors of Eqs.~(\ref{eq:prior_bphi}) and (\ref{eq:prior_bphidelta}), and the remaining are for narrower and wider priors as labeled; all priors are centered around the fiducial bias relations. The grey band in the lower panel simply extends the solid black error bar to ease the comparison, and the dashed grey line marks the fiducial value of $\fnl=5$. The dashed black error bar shows the result obtained with the power spectrum data alone to illustrate the gains from adding bispectrum information; all other results are for the combined power spectrum and bispectrum data.}
\label{fig:param1}
\end{figure}

The marginalized $1\sigma$ constraints on $\fnl$ are shown in the lower panel of Fig.~\ref{fig:param1} for parametrization 1. The upper panels show the two-dimensional $1\sigma$ marginalized constraints on the $\bphi^{\rm A}-\fnl$ and $\bphidelta^{\rm A}-\fnl$ planes; we do not show the contours for all of the 13 parameters for brevity (the interested reader can find some of these triangle plots in Ref.~\cite{2020JCAP...12..031B}). The different colors show the result for different assumed priors on the $\bphi(b_1)$ and $\bphidelta(b_1)$ relations, as labeled; the ``Gaussian prior'' result in orange corresponds to the use of Eqs.~(\ref{eq:prior_bphi}) and (\ref{eq:prior_bphidelta}) as priors.

Assuming perfect knowledge of the bias relations, our fictitious galaxy sample would be able to constrain $\fnl$ with a $1\sigma$ uncertainty of $\sigma_{\fnl} = 1.7$ (solid black error bar). When we adopt our IllustrisTNG-inspired priors for the bias relations of $M_*$-selected galaxies, which recall are centered around the fiducial relations and assume an uncertainty on $\bphi(b_1)$ and $\bphidelta(b_1)$ of $1$ and $5$, respectively (cf.~Eqs.~(\ref{eq:prior_bphi}) and (\ref{eq:prior_bphidelta})), then the $\fnl$ constraint becomes $\fnl = 4.1^{+2.2}_{-1.6} (1\sigma)$ (orange error bar), i.e., the total $1\sigma$ interval increases by about $10\%$ and the mean distribution value gets shifted from the truth by $\approx 0.5\sigma_{\fnl}$. This shift is a manifestation of projection effects that can be understood as follows. In the parts of the parameter space that are close to $\fnl = 0$ (which our chains still explore), the theory predictions become insensitive to both $\bphi$ and $\bphidelta$, which can take on any value allowed by the assumed priors. The wider the priors, the greater the volume of the parameter space that is near the $\fnl = 0$ direction, and so marginalizing over the poorly constrained $\bphi$ and $\bphidelta$ will progressively center the marginalized constraints around $\fnl = 0$. Indeed, the green and blue error bars show the result for $2\times$ and $10\times$ wider priors, which becomes more biased, as expected. On the other hand, halving the width of the priors (magenta), tightens the allowed range of $\bphi$ and $\bphidelta$ sufficiently to the point where the constraints become effectively the same as the ``perfect knowledge'' case. A cautionary tale here is that contrary to what one might have naively expected, wide priors on the bias parameters are not conservative and should be interpreted carefuly in light of projection effects like these.\footnote{We note also for completeness that the constraints on the $\bphi-\fnl$ and $\bphidelta-\fnl$ planes are bimodal because the goodness-of-fit does not vary dramatically under a simultaneous change of sign of $\bphi$, $\bphidelta$ and $\fnl$. This is not visible in the scale of the upper panels in Fig.~\ref{fig:param1} for the $1\sigma$ contours, but see Figs.~2 and 3 in Ref.~\cite{2020JCAP...12..031B} for a clear visualization of the bimodality of the constraints (the upper right panel of Fig.~\ref{fig:param23} in this paper also displays this bimodal nature).}

\begin{figure}
\centering
\includegraphics[width=\textwidth]{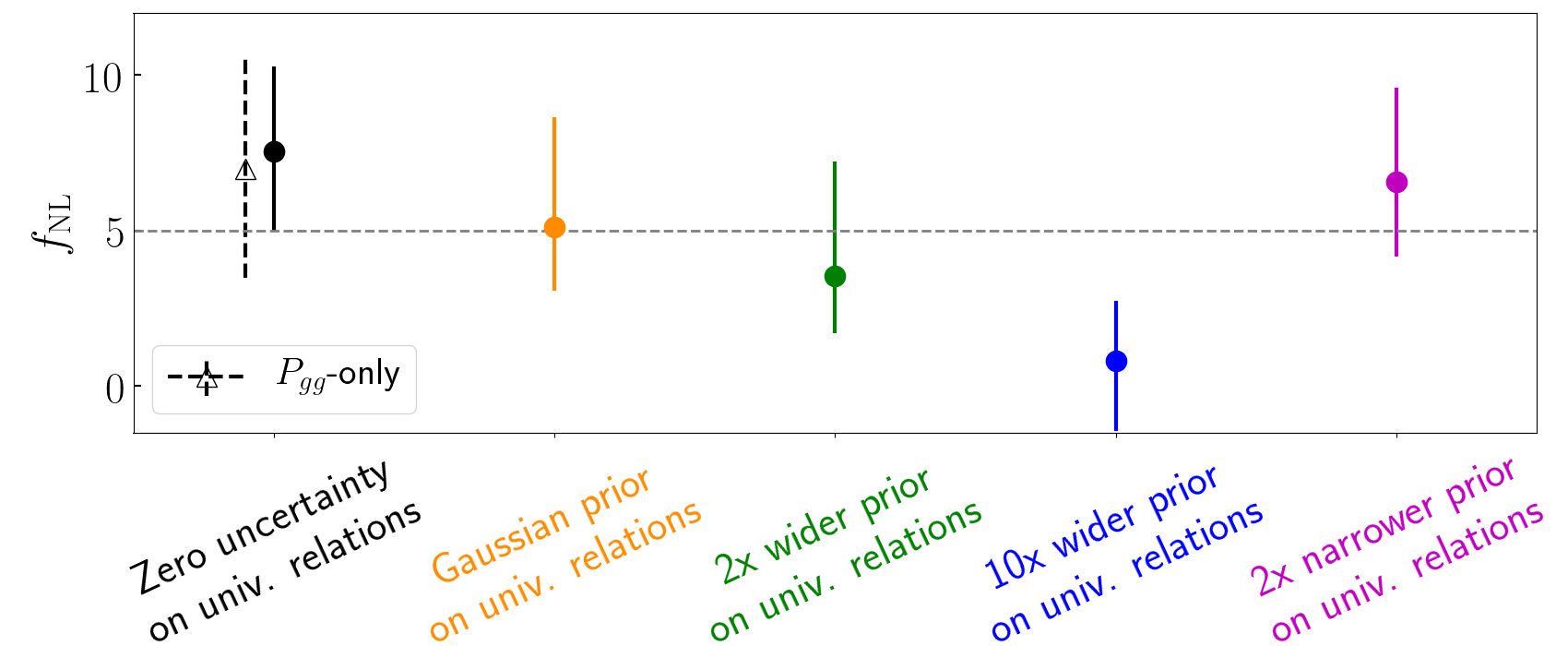}
\caption{Same as the lower panel of Fig.~\ref{fig:param1}, but for priors on $\bphi(b_1)$ and $\bphidelta(b_1)$ that are offset from the fiducial ones. The result in black assumes the universality relations (Eqs.~(\ref{eq:bphi_uni}) and (\ref{eq:bphidelta_uni})) with zero uncertainty, the result in orange is for the Gaussian priors of Eqs.~(\ref{eq:prior_bphi}) and (\ref{eq:prior_bphidelta}) with the mean relations replaced by the universality relations, and the remaining data points are for narrower and wider priors as labeled.}
\label{fig:param1offset}
\end{figure}

In real-life applications, however, the assumed priors will likely not be exactly centered around the bias of the observed galaxies, which introduces additional shifts in the best-fitting $\fnl$ and its inferred error bar. This is illustrated in Fig.~\ref{fig:param1offset}, which shows the same as the lower panel of Fig.~\ref{fig:param1}, but for offset priors on the bias parameters centered around the universality relations, instead of the fiducial relations used to generate the data vector. Assuming zero uncertainty on the wrong bias relations (black error bar), our fictitious survey would constrain $\fnl = 7.5 \pm 2.5\ (1\sigma)$, i.e., the inferred value would be $1\sigma$ away from the fiducial $\fnl=5$ (the upward shift in the constraint and the increase of the size of the error bar compared to Fig.~\ref{fig:param1} is as expected since the universality relations underpredict the fiducial bias values). {If one is interested only on the overall detection of $f_{\rm NL} \neq 0$, then this represents a $\approx 3\sigma$ detection, which is the same significance as the constraint $\fnl = 5.0 \pm 1.7\ (1\sigma)$ in Fig.~\ref{fig:param1} assuming perfect knowledge of the fiducial bias relations. Thus, if the true value is $\fnl \neq 0$, then from the point of view of {\it detection significance}, the assumptions on the bias parameters are not as critical. Note, however, that different assumptions on the bias relations still directly impact the inferred precision of the constraint $\sigma_{\fnl}$, which can lead to misinterpretations about the true constraining power of the data. For detection significance purposes, a much cleaner approach is thus that based on parametrization 2 discussed below.} Further, as the width of the prior increases in Fig.~\ref{fig:param1offset}, the constraint progressively approaches $\fnl=0$ due to the projection effects discussed above. Interestingly, the result in orange shows the case for widths of $\Delta\bphi = 1$ and $\Delta\bphidelta=5$, for which the projection effects balance the shifts induced by centering the prior around the wrong $\bphi(b_1)$ and $\bphidelta(b_1)$ relations.

{Although not shown, we have also repeated the analyses behind Figs.~\ref{fig:param1} and \ref{fig:param1offset}, but for a fiducial value of $\fnl = 0$, instead of $\fnl = 5$. In this case, for both the cases with priors centered around and offset from the fiducial bias relations, the constraints on $\fnl$ are always unbiased, but too wide priors on $\bphi(b_1)$ and $\bphidelta(b_1)$ eventually also artificially shrink the marginalized error bar due to the projection effects. For the case of assuming perfect knowledge of the bias relations and zero uncertainty around the wrong ones, our fictitious survey would constrain $|\fnl| < 1.2\ (1\sigma)$ and $|\fnl| < 2.0\ (1\sigma)$, respectively, further illustrating how wrong assumptions about the bias relations can directly impact the apparent constraining power of the data.}

{We emphasize that} the absolute values of the constraints in Figs.~\ref{fig:param1} and \ref{fig:param1offset} correspond strictly to our fictitious galaxy survey, but the relative impact of the galaxy bias uncertainties on those constraints is a more trustworthy measure of what to expect for real surveys. Taken at face value, our results in Fig.~\ref{fig:param1} suggest that priors on the $\bphi(b_1)$ and $\bphidelta(b_1)$ relations centered around the truth with widths of order $0.5-1$ and $2.5-5$, respectively (i.e., in between the results shown in orange and magenta in Fig.~\ref{fig:param1}), may be necessary to guarantee unbiased constraints on $\fnl$. Within the IllustrisTNG model, Fig.~\ref{fig:bphi_bphidelta_all} suggests that this target might be achievable for the case of $M*$-selected galaxies with more simulations to beat the statistical errors. However, the figure shows that the $\bphi(b_1)$ and $\bphidelta(b_1)$ relations for other selection criteria differ by more than these target widths, and it is still currently unclear how sensitive the bias relations are to changes in the star formation and baryonic feedback models in galaxy formation simulations. Further, recall also the lesson from Fig.~\ref{fig:param1offset} that considerations about the width of the priors on the bias relations cannot be disentangled from discussions about the assumed center values. This all strongly motivates more work to study the bias relations in galaxy formation models beyond IllustrisTNG, as well as for more refined criteria to select the objects, in particular, criteria that resemble as closely as possible those adopted for real galaxies. We highlight that progress along these lines may prove critical to the success of future galaxy surveys to improve over the CMB constraints on local PNG.

\subsection{Results from parametrization 2:  fit for products $\fnl b_\phi$, $\fnl \bphidelta$}
\label{sec:forecasts_param2}

\begin{figure}
\centering
\includegraphics[width=\textwidth]{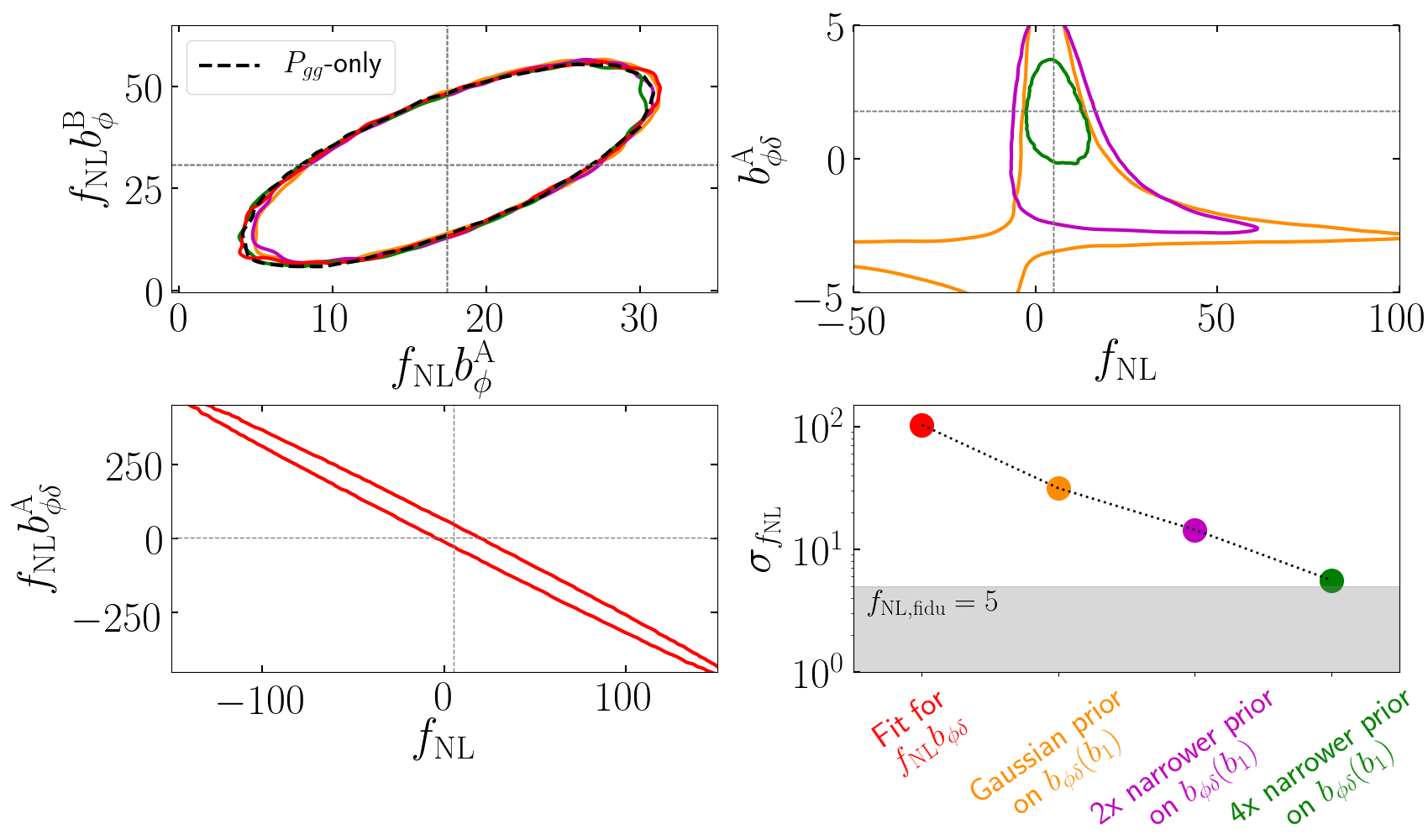}
\caption{Constraint results from parametrization 2 that fits directly for products $\fnl b_\phi$, $\fnl \bphidelta$. The upper panels and the lower left panel show two-dimensional marginalized $1\sigma$ constraints, whereas the lower right panel shows the marginalized $1\sigma$ errors on the parameter $\fnl$ as a function of different assumed priors on $\bphidelta$, as labeled. The results in red correspond to fitting for $\fnl \bphidelta^{\rm A}$ and do not require priors on $\bphidelta$. All results are for the combined power spectrum and bispectrum data, except that in dashed black in the upper left panel which is only for the power spectrum data.}
\label{fig:param23}
\end{figure}

Our current limited understanding of the $\bphi(b_1)$ and $\bphidelta(b_1)$ relations motivates the exploration of constraints on local PNG using parametrization 2, which constrains the products $\fnl b_\phi$ and $\fnl \bphidelta$, and avoids the need to put priors on the local PNG bias parameters. For our forecast setup, the overall significance of the detection of local PNG is then determined by the combined significance of the detection of nonzero values for the parameters $\fnl b_\phi^{\rm A}$, $\fnl b_\phi^{\rm B}$, $\fnl$ and $\fnl\bphidelta^{\rm A}$. The results are displayed in Fig.~\ref{fig:param23} in red color. For $\fnl \bphi^{\rm A}$ and $\fnl \bphi^{\rm B}$, the top left panel shows that our fictitious survey would comfortably detect these parameters at the $1\sigma$ level. However, as noted previously in Ref.~\cite{2020JCAP...12..031B}, these constraints are dominated by the power spectrum part of the data vector, without much contribution from the bispectrum. This can be seen by comparing the solid red contour in the top left panel, which is for the combined power spectrum and bispectrum data, with the dashed black contour for the $P_{gg}$-only analysis: the two are barely distinguishable, which shows that the bispectrum constrains the $\fnl \bphi^{\rm A}$ and $\fnl \bphi^{\rm B}$ parameters only weakly. In parametrization 2, the bispectrum thus contributes to the overall detection of local PNG only via a detection of the parameters $\fnl$ or $\fnl \bphidelta$, but as the lower left panel of Fig.~\ref{fig:param23} shows, these are strongly degenerate and poorly constrained. Thus, while parametrization 2 can prove a powerful way to detect local PNG with power spectrum data without making assumptions on galaxy bias, the price to pay is that the bispectrum would not be able to further enhance the significance of the detection. This can be compared with the case of parametrization 1 in Fig.~\ref{fig:param1}, in which the bispectrum helps visibly in improving the bounds on $\fnl$: for the ``perfect knowledge'' case, the uncertainty on $\fnl$ shrinks by about $30\%$ when the bispectrum information is considered (cf.~the two error bars in black in the lower panel of Fig.~\ref{fig:param1}).

As an attempt to ``rescue'' the constraining power of the bispectrum, we also show results where, in parametrization 2, we replace $\fnl \bphidelta^{\rm A}$ with $\bphidelta^{\rm A}$, for which we assume Gaussian priors again. The results are shown in orange, magenta and green in Fig.~\ref{fig:param23} for varying prior widths, as labeled. As expected, the upper left panel shows that tightening the prior on $\bphidelta^{\rm A}$ has no visible impact on the $\fnl \bphi^{\rm A}$, $\fnl \bphi^{\rm B}$ constraints, as they are dominated by the power spectrum data where $\bphidelta$ does not contribute. The main effect is thus the progressive tightening of the $\fnl$ bound for tighter $\bphidelta$ priors, as shown on the right of Fig.~\ref{fig:param23}. Specifically, when we fit for $\fnl\bphidelta$, we find $\sigma_{\fnl} \approx 100$, but for our default Gaussian prior on $\bphidelta$, a prior that is $2\times$ narrower, and another prior that is $4\times$ narrower, the uncertainties become, respectively, $\sigma_{\fnl} \approx 30$, $\sigma_{\fnl} \approx 15$ and $\sigma_{\fnl} \approx 5$. That is, it is only after the prior width on $\bphidelta$ becomes of order unity (green; $5/4 = 1.25 \sim 1$) that the constraints return $\sigma_{\fnl}$ values comparable to our fiducial choice of $\fnl = 5$, or in other words, that the bispectrum begins to add to the overall significance of the detection of local PNG.

Before concluding, we clarify the origin of the peculiar shape of the orange contour in the upper right panel of Fig.~\ref{fig:param23}, namely its sudden widening in the $\fnl$ direction. It is possible to show that the $\propto 2b_1^3\fnl$ and $\propto b_1^2\fnl\bphidelta$ contributions to the galaxy bispectrum (cf.~Eq.~(\ref{eq:bggg_model_NG})) have a similar scale-dependence on the range of scales that have the most constraining power (this is especially the small-scale squeezed bispectrum); this is in fact the reason behind the strong degeneracy between $\fnl$ and $\fnl\bphidelta$ discussed above. If the degeneracy was perfect and if $\bphidelta = -2b_1$, then the bispectrum would become independent of $\fnl$ in parametrization 2, and unable to constrain it. In reality, the degeneracy is not perfect, but as the chains approach $\bphidelta^{\rm A} = -2b_1^{\rm A} \approx -3.16$ (which is where the widening occurs), the sensitivity to $\fnl$ still gets significantly reduced, and the contours get wider.

\section{Summary and conclusions}
\label{sec:summary}

Observational constraints on $\fnl$ using galaxy power spectrum and bispectrum data are determined largely from contributions $\propto \bphi\fnl$ and $\propto\bphidelta\fnl$, and as a result, competitive constraints on $\fnl$ using these data require priors on the bias parameters $\bphi$ and $\bphidelta$ to be assumed. The most popular approach encountered in the literature (both in real-data constraints and forecasts) involves using the so-called {\it universality relations} of Eqs.~(\ref{eq:bphi_uni}) and (\ref{eq:bphidelta_uni}) (or certain variants thereof) to relate these two parameters to the parameter $b_1$ that can be constrained using the parts of the data where $\fnl$ contributes only weakly. The problem is that these relations are derived for dark matter halos assuming that their mass function is universal, and thus, there is no reason to expect them to hold for real-life tracers of the large-scale structure like galaxies.

In fact, prior to this work, it was already known that the universality relation does not describe perfectly the $\bphi(b_1)$ relation of halos in gravity-only simulations, nor the relation for stellar-mass selected galaxies in hydrodynamical simulations. In this paper, we tested, for the first time with dedicated measurements from separate universe $N$-body simulations, the validity of the universality relation for the $\bphidelta$ parameter as well, which enters at leading order in the galaxy bispectrum. We carried out our analysis for both dark matter halos in gravity-only simulations, as well as simulated galaxies in hydrodynamical simulations with the IllustrisTNG galaxy formation model. We studied in particular the sensitivity of the $\bphi(b_1)$ and $\bphidelta(b_1)$ relations to different ways of selecting the galaxy samples, including in terms of total mass, stellar mass, black hole mass, black hole mass accretion rate and color. Using an idealized forecast setup with galaxy power spectrum (multitracer) and bispectrum data for a fictitious survey with $V = 100{\rm Gpc}^3/h^3$ at $z=1$, and a fiducial value of $\fnl = 5$, we then explored ways for how to incorporate uncertainties on the $\bphi(b_1)$ and $\bphidelta(b_1)$ relations in $\fnl$ constraint analyses.

Our main results can be summarized as follows:
\begin{itemize}

\item The $\bphidelta(b_1)$ relation of dark matter halos is approximately redshift-independent, but it is not adequately described by the universality relation. The most notable difference is an overprediction by the latter in the range $1 \lesssim b_1 \lesssim 3$, by up to $\Delta\bphidelta \approx 3$ (cf.~Fig.~\ref{fig:bphidelta_b1}). 

\item For the simulated galaxies, the $\bphi(b_1)$ and $\bphidelta(b_1)$ relations are both very sensitive to the galaxy selection criteria, are poorly described by the corresponding universality relations, and do not generically admit a regular, redshift-independent function of $b_1$ (cf.~Fig.~\ref{fig:bphi_bphidelta_all}).

\item Priors on $\bphi(b_1)$ and $\bphidelta(b_1)$ centered close to the fiducial with widths of order $1$ and $5$, respectively, proved sufficient to yield relatively unbiased $\fnl$ constraints (cf.~Figs.~\ref{fig:param1} and \ref{fig:param1offset}). This sets a rough target for how precisely these relations may need to be determined from simulations.

\item Fitting for products of $\fnl\bphi$ and $\fnl\bphidelta$ still allows to rule out $\fnl =0$ without assumptions on galaxy bias, but renders the bispectrum less useful in the constraints. For our fictitious survey analysis, the bispectrum only begins improving the overall detection of local PNG if priors on $\bphidelta(b_1)$ of order $\lesssim 1$ are assumed (cf.~Fig.~\ref{fig:param23}).

\end{itemize}
Overall, our current poor understanding of the $\bphi$ and $\bphidelta$ parameters strongly motivates more works like this one to pin down the expected values of these parameters for real-life galaxy samples. For example, it is important to determine the extent to which the $\bphi(b_1)$ and $\bphidelta(b_1)$ relations depend on the assumed galaxy formation physics. This can be done with simulations for variants of the IllustrisTNG model parameters, or for different self-consistent models of galaxy formation altogether. Given the sensitivity to the galaxy selection criteria that we encountered in this paper, it is also important that future works with galaxy formation simulations attempt also to mimic as closely as possible the selection strategy applied to the real-life galaxy samples considered. Further, beyond galaxies as tracers, it would be interesting to study also the $\bphi$ and $\bphidelta$ parameters of the gas distribution \cite{2020arXiv201204637V}, in particular, the distribution of neutral hydrogen mapped with 21cm line intensity mapping observations.

{It is important to acknowledge, however, that the field of cosmological hydrodynamical simulations of galaxy formation has still many associated uncertainties, and that it can be nontrivial to work out the selection functions of real-life galaxies and apply them on simulated ones. This suggests that obtaining very tight theoretical priors on the $\bphi(b_1)$ and $\bphidelta(b_1)$ relations can prove challenging, but this is work that needs to be carried out anyway to help shape our approach to $\fnl$ constraints using large-scale structure data. For example, even if at the end of the day the conclusion is that our priors on these bias relations are not satisfactory, this will still be informative in that it will suggest abandoning approaches that focus on the numerical value of $\fnl$ (like in parametrization 1), and limiting ourselves to analyses based on parametrization 2 more focused on simply detecting $\fnl \neq 0$.}

We finish by highlighting also the importance for future studies on $\fnl$ to begin taking galaxy bias uncertainties into account. The strategies that we described in Sec.~\ref{sec:forecasts} are straightforward to implement in real-data constraint analyses like those of Refs.~\cite{2019JCAP...09..010C, 2021arXiv210613725M} using eBOSS survey data, as well as in the many forecast codes that exist in the literature for surveys like Euclid \cite{2011arXiv1110.3193L}, SphereX \cite{2014arXiv1412.4872D} or SKA \cite{2020PASA...37....2W}. Doing so is important not only to guarantee robust and trustworthy bounds on $\fnl$, but also to determine more precisely the theoretical precision requirements on the $\bphi$ and $\bphidelta$ parameters for future galaxy surveys to reach their $\fnl$ targets.

\acknowledgments
We would like to thank Giovanni Cabass, Dragan Huterer, Eiichiro Komatsu, Titouan Lazeyras and Fabian Schmidt for very useful comments and discussions. The author acknowledges support from the Excellence Cluster ORIGINS which is funded by the Deutsche Forschungsgemeinschaft (DFG, German Research Foundation) under Germany's Excellence Strategy - EXC-2094-390783311. The numerical analysis of the simulation data presented in this work was done on the Cobra supercomputer at the Max Planck Computing and Data Facility (MPCDF) in Garching near Munich. 


\appendix

\section{Validation of the method to estimate $\bphidelta$ using the $b_2$ parameter}
\label{app:b2}

In this appendix we estimate the bias parameter $b_2$ using a similar method to that used to estimate the $\bphidelta$ parameter. This allows us to assess the performance of our method by comparing against known results for the $b_2(b_1)$ relation of dark matter halos in the literature. Up to a subtle caveat that we comment on below, the method described here is in all similar to that used also by Ref.~\cite{baldauf/etal:2015}.

Starting again from the galaxy-matter cross-power spectrum described by $P_{gm} = b_1P_{mm}$, and defining its linear response to mass perturbations $\delta_m$ as $R_{1, gm} = \partial {\rm ln} P_{gm}/\partial\delta_m$, we have
\bq\label{eq:R1gm}
R_{1, gm}(k) = \frac{\partial{\rm ln}b_1}{\partial\delta_m} + R_{1, mm}(k)  = \frac{b_2}{b_1} - b_1 + R_{1, mm}(k),
\eq
where $R_{1, mm} = \partial {\rm ln} P_{mm}/\partial\delta_m$ is the linear response of the matter power spectrum, and in the second equality we have used that $b_2 = (\partial^2 \bar{n}_g/\partial\delta_m^2)/\bar{n}_g$. Thus, given measurements of the power spectrum responses from separate universe simulations and estimates of the values of $b_1$, we can use the above equation to fit for $b_2$.  Alternatively, and as we did in our main results, we can estimate $b_2$ with higher signal-to-noise by computing the response of $b_1$ measured in the separate universe simulations, i.e., using directly
\bq\label{eq:b_2app}
b_2 = \bigg[\frac{\partial{\rm ln}b_1}{\partial\delta_m} + b_1 \bigg] b_1,
\eq
where
\bq\label{eq:respb1app}
\frac{\partial{\rm ln}b_1}{\partial\delta_m} = \frac{1}{2}\left[\frac{\partial{\rm ln}b_1}{\partial\delta_m}\right]^{{\rm High}\delta_m} + \frac{1}{2}\left[\frac{\partial{\rm ln}b_1}{\partial\delta_m}\right]^{{\rm Low}\delta_m},
\eq
with
\bq\label{eq:respb1_2app}
\left[\frac{\partial{\rm ln}b_1}{\partial\delta_m}\right]^{{\rm High} \delta_m} &=& \frac{1}{\delta^{\rm High \delta_m}_L(z)}\Big[\frac{b_1^{\rm High \delta_m}}{b_1^{\rm Fiducial}} - 1\Big], \nonumber \\
\left[\frac{\partial{\rm ln}b_1}{\partial\delta_m}\right]^{{\rm Low} \delta_m} &=& \frac{1}{\delta^{\rm Low \delta_m}_L(z)}\Big[\frac{b_1^{\rm Low \delta_m}}{b_1^{\rm Fiducial}} - 1\Big].
\eq
In these equations, $\delta^{\rm High \delta_m}_L(z)$ and $\delta^{\rm Low \delta_m}_L(z)$ are the amplitudes of the matter perturbation in the separate universe cosmologies, and the superscripts indicate in which cosmology $b_1$ is evaluated using Eq.~(\ref{eq:b1_method}). We evaluate the above equations using the separate universe simulations presented in Refs.~\cite{2019MNRAS.488.2079B, 2020JCAP...12..013B}, which correspond to the same Fiducial cosmology used in the main body of the paper, and two separate universe cosmologies, ${\rm High} \delta_m$ and ${\rm Low} \delta_m$, characterized by $\delta^{\rm High \delta_m}_L(z=0) = +0.05$ and $\delta^{\rm Low \delta_m}_L(z=0) = -0.05$, respectively. 

\begin{figure}
\centering
\includegraphics[width=\textwidth]{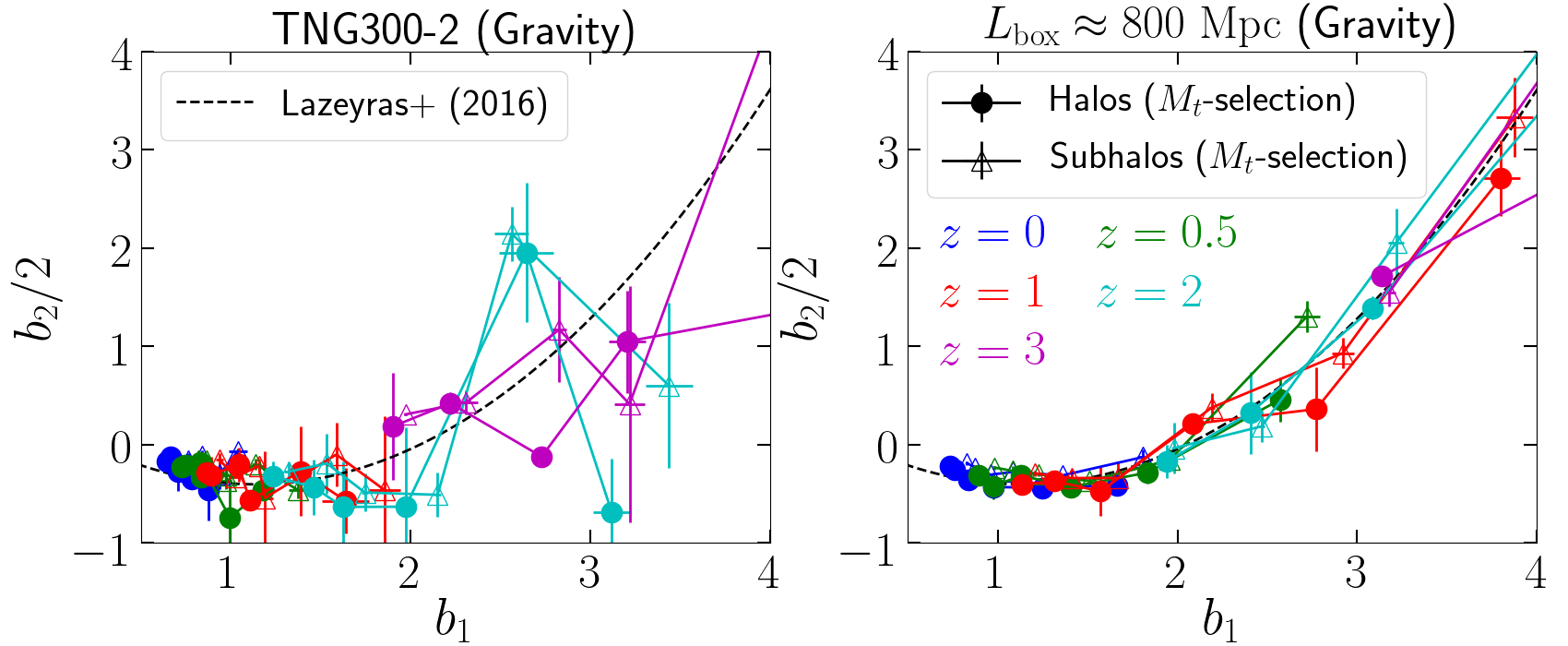}
\caption{The $b_2(b_1)$ relation for halos (filled symbols) and subhalos (open symbols) in the gravity-only simulations. This is the same as Fig.~\ref{fig:bphidelta_b1} in the main text, but for $b_2$, instead of $\bphidelta$. In both panels, the dashed line shows the fit obtained by Ref.~\cite{lazeyras/etal} for dark matter halos using separate universe simulations (cf.~Eq.~(\ref{eq:b2titouan})).}
\label{fig:b2_b1}
\end{figure}

The caveat that is important to mention here concerns the choice of the box size in these separate universe simulations. Contrary to the changes in $\A_s$ that characterize the ${\rm High}\A_s$ and ${\rm Low}\A_s$ cosmologies in the main body of the paper, the values of $\delta_L(z)$ modify the time-evolution of the scale-factor $a(t)$, and consequently, the relationship between comoving and physical distances. The separate universe simulations in Refs.~\cite{2019MNRAS.488.2079B, 2020JCAP...12..013B} are for fixed comoving size of the boxes at all epochs in units of $\rm Mpc$. This is important for how to interpret the corresponding response measurements since the finite-differences evaluated with these simulations are then being taken at fixed comoving volume, but strictly, the full responses in the equations above include the effects from changing the volume. For the case of the power spectrum responses, this means that the measured response corresponds actually to the so-called {\it growth-only} piece. Explicitly, the power spectrum response functions can be decomposed as \cite{li/hu/takada, response}
\bq\label{eq:Rdec}
R_{1, mm}(k) &=& 1 - \frac{1}{3}\frac{{\rm dln}P_{mm}(k)}{{\rm dln}k} + G_{1, mm}(k), \\
R_{1, gm}(k) &=& - \frac{1}{3}\frac{{\rm dln}P_{gm}(k)}{{\rm dln}k} + G_{1, gm}(k),
\eq
where $G_{1, mm}$ and $G_{1, gm}$ are the growth-only responses that one measures with fixed-comoving-volume separate universe simulations, and the remaining terms account for so-called {\it reference density} and {\it dilation} effects, that describe respectively, the fact that in the separate universe cosmology the power spectrum is measured w.r.t.~a modified mean density, and that the physical scales get modified (or {\it diluted}) by the modified scale factor; the missing $+1$ term in $R_{1, gm}$ has to do with the fact that there is one less instance of $\delta_m$ in $P_{gm}$ and the galaxy overdensity is typically measured with respect to the observed local, mean galaxy density (and not the global one). Plugging these expressions in Eq.~(\ref{eq:R1gm}) gives (noting that on large scales the two logarithmic derivative terms are the same)
\bq\label{eq:R1gm_2}
G_{1, gm}(k) = \frac{b_2}{b_1} - b_1 + 1 + G_{1, mm}(k) \equiv \left.\frac{\partial{\rm ln}b_1}{\partial\delta_m}\right\vert_{\rm fixed\ vol.} + G_{1, mm}(k),
\eq
where the second equality defines the fixed-volume derivative of $b_1$ that we actually evaluate with our separate universe simulations using Eqs.~(\ref{eq:respb1app}) and (\ref{eq:respb1_2app}), and which is related to the full, physical derivative as
\bq\label{eq:twoderivs}
\left.\frac{\partial{\rm ln}b_1}{\partial\delta_m}\right\vert_{\rm fixed\ vol.} = \frac{\partial{\rm ln}b_1}{\partial\delta_m} + 1.
\eq
We thus estimate $b_2$ using
\bq\label{eq:b_2app_2}
b_2 = \Bigg[\left.\frac{\partial{\rm ln}b_1}{\partial\delta_m}\right\vert_{\rm fixed\ vol.} + b_1 - 1 \Bigg] b_1.
\eq
The result is in Fig.~\ref{fig:b2_b1}, which shows the same as Fig.~\ref{fig:bphidelta_b1}, but for $b_2$ instead of $\bphidelta$. Our estimated $b_2(b_1)$ relation recovers well the expected result shown by the dashed line, obtained by Ref.~\cite{lazeyras/etal} for dark matter halos using separate universe simulations (cf.~Eq.~(\ref{eq:b2titouan})). Likewise for $\bphidelta$, the results have higher signal-to-noise in the $\bigbox$, compared to the TNG300-2 box. Overall, this successful recovery of the expected $b_2(b_1)$ relation validates our estimates of $\bphidelta(b_1)$ in the main body of the paper, which were obtained effectively in the same manner.

\section{Expressions of the tree-level galaxy power spectrum and bispectrum}
\label{app:theory}

In Sec.~\ref{sec:forecasts}, we evaluate the galaxy power spectrum part of the data vector as (keeping only terms that are leading-order in $\fnl$)
\bq
\label{eq:pgg_modelAA}P^{\rm AA}_{gg}(k) &=& \left[b_1^{\rm A}\right]^2 P_{mm}(k) + 2b_1^{\rm A}b_\phi^{\rm A} \fnl P_{m\phi}(k) + P^{\rm AA}_{\eps\eps}, \\
\label{eq:pgg_modelAB}P^{\rm AB}_{gg}(k) &=& b_1^{\rm A}b_1^{\rm B} P_{mm}(k) + \left[b_1^{\rm A}b_\phi^{\rm B} + b_1^{\rm B}b_\phi^{\rm A}\right] \fnl P_{m\phi}(k), \\
\label{eq:pgg_modelBB}P^{\rm BB}_{gg}(k) &=& \left[b_1^{\rm B}\right]^2 P_{mm}(k) + 2b_1^{\rm B}b_\phi^{\rm B} \fnl P_{m\phi}(k) + P^{\rm BB}_{\eps\eps},
\eq
and the bispectrum part as
\bq\label{eq:bggg_model}
B_{ggg}(k_1,k_2,k_3) = B_{ggg}^{\rm G}(k_1,k_2,k_3) + B_{ggg}^{\rm NG}(k_1,k_2,k_3),
\eq
with the $\fnl$-independent part given by
\bq\label{eq:bggg_model_G}
&&B_{ggg}^{\rm G}(k_1,k_2,k_3) = b_1^3B_{mmm}(k_1, k_2, k_3) + \big[2 b_1 P_{mm}(k_1)P_{\eps\eps_\delta} + {\rm (2\ perm.)}\big] + B_{\eps\eps\eps} \nonumber \\
&+&\Big[b_1^2b_2P_{mm}(k_1)P_{mm}(k_2) + {\rm (2\ perm.)}\Big] + \Big[2b_1^2b_{K^2}\left(\mu_{12}^2 - \frac13\right)P_{mm}(k_1)P_{mm}(k_2) + {\rm (2\ perm.)}\Big] \nonumber \\
\eq
and the $\propto \fnl$ part by
\bq\label{eq:bggg_model_NG}
&&B_{ggg}^{\rm NG}(k_1,k_2,k_3) = \Bigg[ 2b_1^3\fnl\frac{P_{mm}(k_1)P_{mm}(k_2)}{\mathcal{M}(k_1)\mathcal{M}(k_2)}\mathcal{M}(k_3) + 2b_\phi\fnl\frac{P_{mm}(k_1)}{\mathcal{M}(k_1)}P_{\eps\eps_\delta} \nonumber \\
&+&  b_1^2b_\phi\fnl P_{mm}(k_1)P_{mm}(k_2) \Bigg(\mu_{12}\bigg(\frac{k_1}{k_2\mathcal{M}(k_1)} + \frac{k_2}{k_1\mathcal{M}(k_2)}\bigg) + 2F_2(k_1, k_2, \mu_{12})\bigg(\frac{1}{\mathcal{M}(k_1)} + \frac{1}{\mathcal{M}(k_2)}\bigg)\Bigg) \nonumber \\
&+&  \bigg(b_1^2b_{\phi\delta}  + b_1b_2b_{\phi} + 2b_1b_{K^2}b_{\phi} \Big(\mu_{12}^2 - \frac13\Big)\bigg)\fnl P_{mm}(k_1)P_{mm}(k_2) \bigg(\frac{1}{\mathcal{M}(k_1)} + \frac{1}{\mathcal{M}(k_2)}\bigg) + {\rm (2\ perm.) \Bigg]}. \nonumber \\
\eq
In the above expressions, $\mathcal{M}(k) = (2/3) k^2 T_m(k) / (\Omega_{m0} H_0^2)$ with $T_m$ the matter transfer function, $B_{mmm}(k_1, k_2, k_3) = 2F_2(\vk_1, \vk_2) P_{mm}(k_1)P_{mm}(k_2) + {\rm (2\ perm.)}$ is the matter bispectrum, $F_2(k_1, k_2, \mu_{12}) = 5/7 + ({\mu_{12}}/{2})\left[{k_1}/{k_2} + {k_2}/{k_1}\right] + (2/7) \mu_{12}^2$ is the second-order mode-coupling kernel in perturbation theory \cite{Bernardeau/etal:2002}, $\mu_{ab}$ is the cosine angle between the wavenumbers $k_a$ and $k_b$ in the sides of the bispectrum triangle, and $P_{m\phi}(k) = P_{mm}(k) / \mathcal{M}(k)$ is the matter-potential cross-power spectrum. We calculate all of the spectra and transfer functions with the {\tt CAMB} code \cite{camb}. Further, $P_{\eps\eps}$,  $P_{\eps\eps_\delta}$ and $B_{\eps\eps\eps}$ are the power spectra and bispectra of the stochastic terms.

In our analysis, we consider $k_{\rm max} = 0.2h/{\rm Mpc}$, on which corrections to our tree-level theory model do become important, but since the data vector is obtained from the theory model this does not impact our main conclusions on the impact of galaxy bias uncertainties. For simplicity, we also skip modeling the effects of redshift space distortions \cite{2016JCAP...06..014T}, so-called projection/relativistic effects \cite{2011JCAP...10..031B, 2010PhRvD..82h3508Y, challinor/lewis:2011, gaugePk, 2016JCAP...05..009R, 2015ApJ...814..145A, 2020arXiv200701802W, 2021JCAP...04..013M}, and observational systematics \cite{2021arXiv210613724R}. Any reduction in constraining power from taking these complications into account reduces the overall importance of the uncertainties on galaxy bias, but only in the sense that there are more sources of uncertainty worsening the constraints on $\fnl$. Our results in the main body of the paper thus assess how well we need to understand the $\bphi$ and $\bphidelta$ parameters, assuming negligible sources of error from other aspects of observational constraints on local PNG.

The covariance matrix of the data vector can be written as
\begin{equation}\label{eq:cov_def}
{\Cov} = 
\begin{pmatrix}
\Cov^{PP}  & \Cov^{BP} \\[1.5ex]
\cdots  & \Cov^{BB} \\[1.5ex]
\end{pmatrix}
\,\,,
\end{equation}
where $\cov^{PP}$, $\cov^{BP}$ and $\cov^{BB}$ indicate the covariance of the power spectrum part of the data vector, the cross-covariance of the bispectrum and power spectrum, and the covariance of the bispectrum part, respectively. For $\cov^{PP}$ we consider only the {\it Gaussian+shot noise} contribution, for $\cov^{BB}$ we consider in addition to the Gaussian contribution also contributions from leading-order non-Gaussian terms, and we set $\cov^{BP} = 0$ {for simplicity}. We do not repeat our covariance expressions here, but the interested reader can find them in Ref.~\cite{2020JCAP...12..031B}; see also App.~B there for a look into the impact of different covariance compositions {(including the cross-covariance term $\cov^{BP} = 0$)} on the final $\fnl$ bounds, which is another aspect of these analyses that is often overlooked too.

\bibliographystyle{utphys}
\bibliography{REFS}

\end{document}